\documentclass[%
 reprint,
 amsmath,amssymb,
 aps,
]{revtex4-2}

\usepackage{graphicx}
\usepackage{dcolumn}
\usepackage{bm}
\usepackage{amsmath}
\usepackage{mathrsfs}
\usepackage{xcolor} 
\usepackage[
    colorlinks=true,
    linkcolor=blue,
    citecolor=blue,
    urlcolor=blue
]{hyperref}

\begin{document}

\preprint{APS/123-QED}

\title{Charged Higgs Search at Future Neutrino Telescope and Higgs Factory}

\author{YiCheng Dai}
\email{YichengDai@mail.ecust.edu.cn}
\affiliation{
School of Physics, East China University of Science and Technology, 130 Meilong Road, Shanghai 200237, P. R. China
}

\author{Wei Liao}
\email{liaow@ecust.edu.cn}
\affiliation{
School of Physics, East China University of Science and Technology, 130 Meilong Road, Shanghai 200237, P. R. China
}

\author{Yi-Song Lu}
\email{luys5@mail2.sysu.edu.cn}
\affiliation{
School of Physics and Astronomy, Sun Yat-sen University, Zhuhai 519082, China
}

\author{Qi-Shu Yan}
\affiliation{
School of Physics Sciences, University of Chinese Academy of Sciences, Beijing 100049, P. R. China
}
\affiliation{
Center for Future High Energy Physics, Chinese Academy of Sciences, Beijing 100049, P. R. China
}

\date{\today}

\begin{abstract}
We investigate the discovery potential of the charged Higgs boson at future lepton colliders and neutrino telescopes 
within two simplified benchmark scenarios with universal Yukawa couplings. 
We demonstrate that both the muon-track and cascade events at neutrino telescopes can be exploited to search for 
or constrain the charged Higgs boson through resonant neutrino scattering process, 
and we compare the sensitivities obtained under different astrophysical neutrino flux models. 
We study the effect of the detector volume of neutrino telescopes on the discovery potential of the charged Higgs boson.
We study the signal of charged Higgs boson at future lepton colliders over different kinematic regimes using optimized reconstruction strategies. 
We compare the discovery potential of the two experimental approaches over the relevant parameter space and find that future lepton colliders generally provide better sensitivity in most of the parameter space, while neutrino telescopes with very large detector volume can offer competitive and complementary 
sensitivity in the heavy-mass region of the charged Higgs boson.
\end{abstract}

\maketitle


\section{\label{sec:Introduction}Introduction}

The discovery of the Higgs boson at the Large Hadron Collider (LHC)~\cite{ATLAS:2012yve, CMS:2012qbp} established the existence of a scalar sector responsible for electroweak symmetry breaking. 
 An extended Higgs sector~\cite{Branco:2011iw, Ellwanger:2009dp} is a well motivated possibility in many scenarios beyond the Standard Model (SM). A generic prediction of such scenarios is the existence of charged Higgs bosons, whose observation would provide unambiguous evidence for physics beyond the SM. 
Compared with the LHC, future Higgs factories, such as the Circular Electron Positron Collider (CEPC) and the Future Circular Collider (FCC-ee)~\cite{CEPCStudyGroup:2018ghi,FCC:2018evy,cheng2024physicspotentialcepcprepared,agapov2022futurecircularleptoncollider},
which aims at precise measurements of the SM Higgs boson, 
provide a cleaner experimental environment with significantly reduced QCD backgrounds. 
They can also serve as discovery machines for Higgs sector beyond the SM, e.g., the charged Higgs boson.

In addition to the conventional collider experiment, large volume neutrino telescopes can also serve as detectors at energy frontier.
Existing neutrino telescopes have already detected neutrinos events of PeV and higher energies.
IceCube has reported several PeV-scale events~\cite{IceCube:2021rpz,IceCube:2014stg,Schneider:2019ayi}, while KM3Net has recently observed an ultra-high-energy event with an estimated neutrino energy of approximately $220~{\rm PeV}$~\cite{KM3NeT:2025npi}.
Through scattering off electrons at rest, neutrinos in this energy range can probe center-of-mass energies
$\sqrt{s}\simeq\sqrt{2m_eE_\nu}$ from tens to several hundreds GeV.

Several next-generation neutrino telescopes are being constructed or proposed~\cite{Aartsen_2021, CHEN2026171374, ye2024TRIDENT}.
They are designed to achieve detector volumes beyond the cubic-kilometre scale of current neutrino telescopes, such as IceCube~\cite{Aartsen_2017}.
For example, IceCube-Gen2 is designed to provide an optical array of about $8~{\rm km}^3$ in Antarctic ice~\cite{Aartsen_2021}, 
while the proposed deep-sea project High-energy Underwater Neutrino Telescope (HUNT) is designed with detector volume $30~\mathrm{km}^3$~\cite{CHEN2026171374}.
These next generation large volume neutrino telescopes are expected to detect neutrino scattering events with center of mass energies much larger than  those already observed, and are good probes of the physics beyond the SM, e.g., the charged Higgs bosons.

Signals of charged Higgs bosons at IceCube or IceCube-Gen2 experiments have been discussed in some models of
charged Higgs bosons~\cite{Dey_2021, Babu:2019vff, Babu:2022fje,Bai:2025pef}.
In this work, we conduct largely model-independently a detailed study on the observable track and cascade events and
the discovery sensitivity of the charged Higgs boson at future neutrino telescope with very large detector volumes.
We study the effect of the detector volume of neutrino telescopes on the discovery sensitivity of the charged Higgs boson.
We also investigate the sensitivity of the charged Higgs search at future Higgs factories
and compare the discovery potentials of the charged Higgs bosons at both future Higgs factories and future large volume neutrino telescopes.
The rest of the paper is organized as follows. 
In Section.~\ref{sec:IceCube Sensitivity to Resonant Charged Higgs Scattering}, we introduce the charged Higgs model and the astrophysical neutrino flux models used in our study.
We present search sensitivities of charged Higgs bosons at future neutrino telescopes. 
The search sensitivity of the charged Higgs boson at future Higgs factories are analyzed in Section.~\ref{Collider Signal}. 
We summarize our results in Section.~\ref{Summary}.

\section{\label{sec:IceCube Sensitivity to Resonant Charged Higgs Scattering}
Resonant Charged Higgs Scattering at Future Neutrino Telecsope }

Ultra-high energy(UHE) neutrino events at neutrino telescopes can be broadly classified into two classes: cascades and tracks.
Cascade events can be produced by electromagnetic or hadronic showers initiated by electrons, hadrons within the detector.
Track events of high-energy muons produced by  neutrinos incident on target can propagate for several kilometers or more  distance in ice or in water\cite{IceCube:2013dkx, IceCube:2020acn}.
We consider both cascade and muon-track events in our analysis of future neutrino telescopes.
At a neutrino telescope, cascade-like and track-like events have experimentally distinguishable  topologies and thus provide independent probes. 
We therefore consider the track and cascade signatures separately in the following discussions.

\subsection{\label{subsection: Charged Higgs Framework for Resonant Scattering}Charged Higgs Framework for Resonant Scattering}
The relevant Yukawa couplings of the charged Higgs boson in a simple model can be described as
\begin{equation}
\mathcal{L} \supset 
H^+ \left( y_{q_i}\,\bar{u}_{L,i} d_{R,i} + y_{\ell}\,\bar{\nu}_L \ell_R \right) + \text{h.c.},
\end{equation}
where $y_\ell\in\{y_e, y_\mu,y_\tau\}$ and $y_{q_i}(i=1,2,3)$ are Yukawa couplings.
The leptonic couplings allow
\begin{equation}
\bar{\nu}_e+e^- \to H^- \to \ell^-+\bar{\nu}_\ell,
\qquad
\ell=e,\mu,\tau,
\end{equation}
while the quark couplings allow the hadronic decay channels
\begin{equation}
\bar{\nu}_e+e^- \to H^- \to d_i+\bar{u}_i,
\qquad
i=1,2,3,
\end{equation}
In order to simplify our discussions, we consider two simplified models
\begin{eqnarray}
&& \textrm{Model I:} ~y_{q_1}=y_{e,\mu}=y,  ~y_{q_2,q_3}=y_\tau=0,\\
&& \textrm{Model II:} ~y_{q_1,q_2}=y_{e,\mu,\tau}=y,  ~y_{q_3}=0.
\end{eqnarray}
The electron and hadronic final states produce cascade events, whereas the muon gives a track event.
$\tau$ lepton produced in the process would soon decay which may give electrons, muon or hadrons in the final state.
Depending on the final states of $\tau$ decay,  it would give cascade or track events at neutrino telescope.
We discuss these events in detail in the following.

\subsubsection{Muon-track signals}
\label{Muon-track signals}
\begin{figure}[t]
    \centering
    \includegraphics[width=0.2\textwidth]{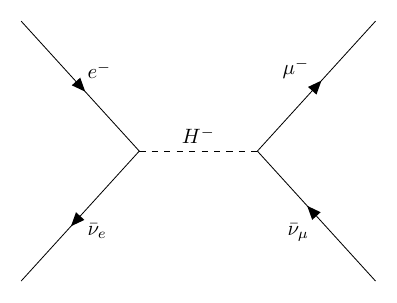} 
    \hspace{0.06\textwidth}
    \includegraphics[width=0.2\textwidth]{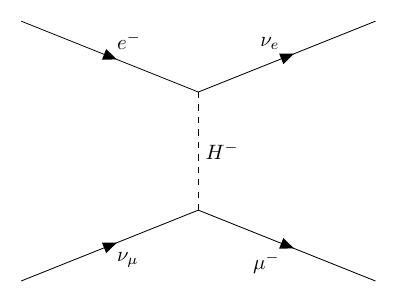}
    \caption{
Feynman diagrams for the charged Higgs induced leptonic processes considered in this section.
The left panel shows the resonant $s$-channel process $\bar{\nu}_e e^- \to H^- \to \mu^- \bar{\nu}_\mu$,
which gives the main charged Higgs signal.
The right panel shows the $t$-channel charged Higgs process
$\nu_\mu e^- \to \mu^- \nu_e$.
}
    \label{fig_s&t}
\end{figure}

The coupling of charged Higgs boson to the leptonic sector allows UHE electron antineutrinos to scatter off electrons in the detector target and produce a muon and a muon antineutrino through an $s$-channel process
\begin{equation}
{\bar \nu}_e + e^- \to H^- \to \mu^- + \bar\nu_\mu .  \label{eq:s-channel}
\end{equation}
These couplings also allow a $t$-channel production process
\begin{equation}
\nu_\mu + e^-\to \mu^- + \nu_e \label{t-channel}
\end{equation}
as shown in Fig.~\ref{fig_s&t}.

 The outgoing high-energy muon leaves a track event in the large volume of the neutrino telescope and can be detected by the Cherenkov detector arrays.
We take these track events without an accompanying cascade as the first signal signature.
The antineutrino-electron cross section of the $s$-channel process shown in Eq.~\eqref{eq:s-channel} can be written as \cite{PhysRev.118.316, Dey_2021, Huang_2020}
\begin{equation}
\begin{aligned}
& \sigma_{H^-}(E_{\bar{\nu}})
=
\frac{16\pi m_e E_{\bar{\nu}}}{m^2_{H^-}}  \\
& \times \frac{
\mathrm{Br}\!\left(H^-\rightarrow e^-\bar{\nu}_e\right)
\mathrm{Br}\!\left(H^-\rightarrow \mu^- \bar{\nu}_\mu \right)
\Gamma^2_{H^-}
}{
\left(2m_eE_{\bar{\nu}}-m_{H^-}^2\right)^2
+
\Gamma_{H^-}^2 m_{H^-}^2
}.
\end{aligned}
\label{eq: Charged_Higgs_cross_section}
\end{equation}
where $E_{\bar{\nu}}$ is the energy of the antineutrino incident on electrons at rest, $\Gamma_{H^-}$ is the total decay width and $m_{H^-}$ and $m_e$ are the masses of the charged Higgs and the electron, respectively.
 In Model I, the charged Higgs boson is assumed to couple universally only to channels of electron, muon and first generation of quarks.
 Therefore, one has $\mathrm{Br}(H^-\to e^-\bar{\nu}_e)=\mathrm{Br}(H^-\to\mu^-\bar{\nu}_\mu)=1/5$, $\mathrm{Br}(H^-\to d\bar u)=3/5$ and
\begin{equation}
\Gamma_{H^-}
=
\frac{m_{H^-}}{16\pi}
\left(y_e^2 + y_\mu^2 + 3y_{q_1}^2\right)
=\frac{5y^2m_{H^-}}{16\pi}
,
\label{eq: Charged_Higgs_width}
\end{equation}
where the coefficient before $y_{q_1}^2$ originates from the color  degree of freedom. 
In Model II, the charged Higgs boson is assumed to couple universally only to the first and second generations of quarks and three generations of leptons, but not to the
third generation of quarks, Therefore, one has $\mathrm{Br}(H^-\to e^-\bar{\nu}_e)=\mathrm{Br}(H^-\to\mu^-\bar{\nu}_\mu)=1/9$, $\mathrm{Br}(H^-\to d\bar u)=\mathrm{Br}(H^-\to s\bar c)=3/9$ and
\begin{equation}
\Gamma_{H^-}
=
\frac{m_{H^-}}{16\pi}
\left(y_e^2 + y_\mu^2 +y_\tau^2+ 3y_{q_1}^2+3y_{q_2}^2\right)
=\frac{9y^2m_{H^-}}{16\pi}
\label{eq: Charged_Higgs_width_Model-II}
\end{equation}

\begin{figure}[t]
    \centering
    \includegraphics[width=0.48\textwidth]{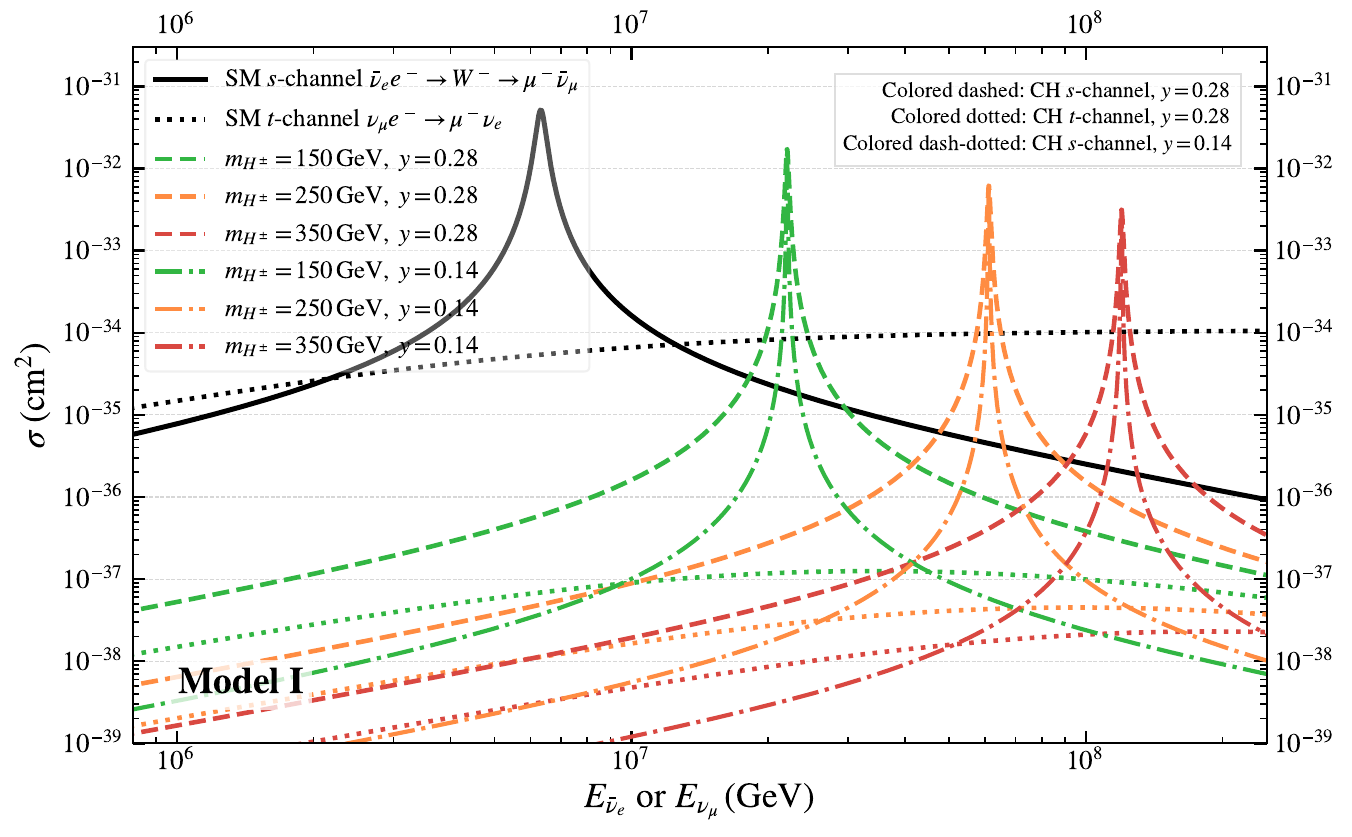}
    \includegraphics[width=0.48\textwidth]{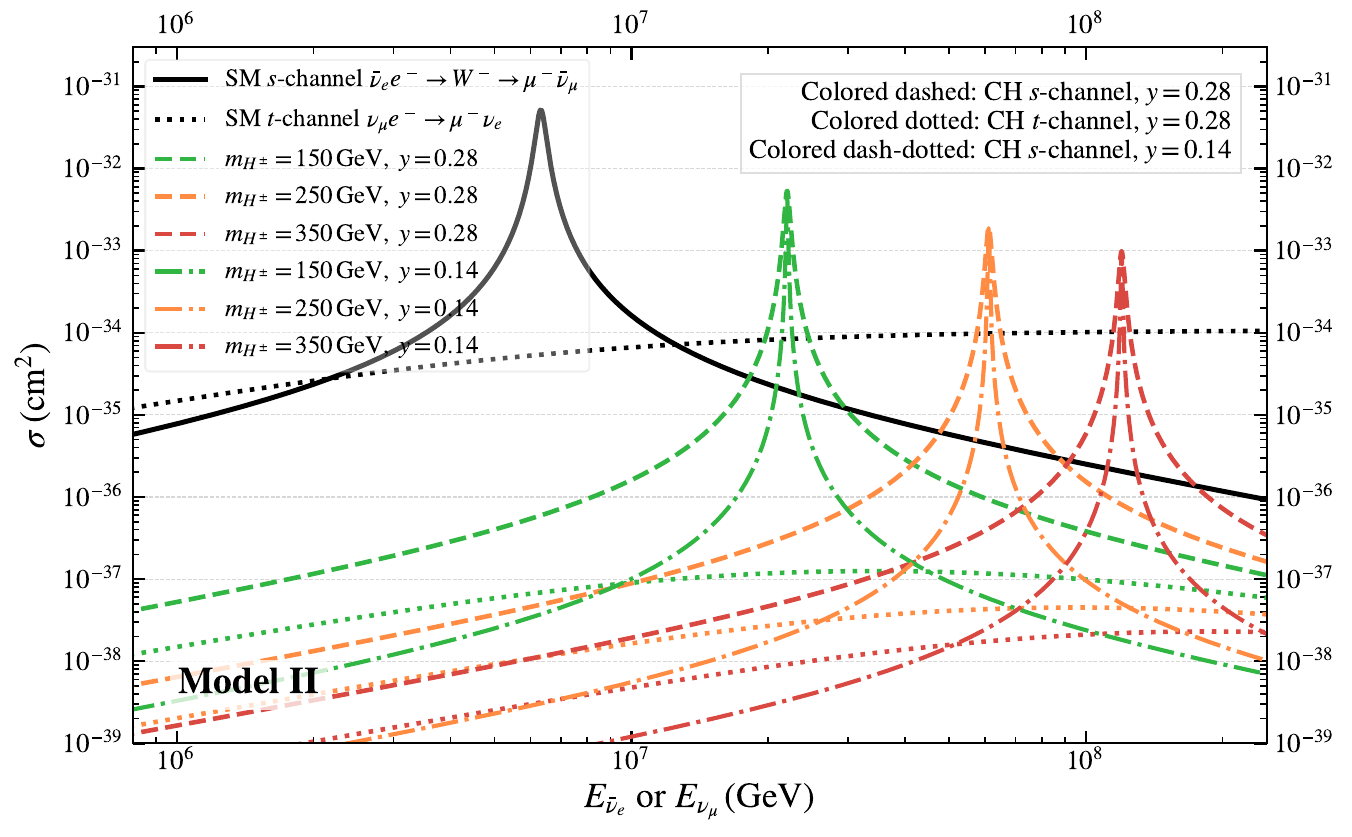}
    \caption{
Comparison of the charged Higgs resonant cross sections and the SM background  for the muon-track channel. 
The black solid curve shows the SM $s$-channel process $\bar{\nu}_e e^- \to W^- \to \mu^- \bar{\nu}_\mu$ 
and the black dotted curve shows the SM $t$-channel W-boson process $\nu_\mu e^- \rightarrow  \mu^-\nu_e$.
The dashed curves show the charged Higgs process $\bar{\nu}_e e^- \to H^- \to \mu^- \bar{\nu}_\mu$ for different charged Higgs masses with a fixed coupling $y=0.28$. 
The dotted colored curves show the corresponding charged Higgs $t$-channel contribution for $y=0.28$. 
The dash-dotted curves show the corresponding charged Higgs results for the same set of charged Higgs masses, but with a smaller coupling $y=0.14$.
The upper and lower panels correspond to Model I and Model II, respectively.
}
    \label{fig:Charged_Higgs_Glashow_cross}
\end{figure}

The incident energy dependence of the resonant cross section for $\bar{\nu}_e e^- \to H^- \to \mu^- \bar{\nu}_\mu$ in Model I is illustrated in the upper panel of Fig.~\ref{fig:Charged_Higgs_Glashow_cross}. 
As shown by the green dash-dotted curve, increasing the coupling $y$ broadens the resonance, as expected from Eq.~(\ref{eq: Charged_Higgs_width}), where $\Gamma_{H^-}\propto y^2$. 
The value of the cross section at the peak of resonance is nearly unchanged because at the resonance position $2m_eE_{\bar{\nu}}=m_{H^-}^2$, the factor $\Gamma_{H^-}^2$ in the numerator of Eq.~(\ref{eq: Charged_Higgs_cross_section}) is canceled by the same factor in the denominator, leaving the peak value only determined by the fixed branching ratios and the resonance mass.
As the mass of the charged Higgs increases, 
the resonance position is shifted to higher energies of incident antineutrino according to
$E_{\rm res}=m_{H^-}^2/(2m_e)$.
At the same time, the value of the cross section at the resonance decreases as \(1/m_{H^-}^2\), as can be seen in  Eq.~\eqref{eq: Charged_Higgs_cross_section}.
For fixed coupling $y$, the width $\Gamma_{H^-}$ increases with $m_{H^-}$.
Nevertheless, for the couplings considered here, the resonance remains narrow, and the resonant cross section of charged Higgs drops rapidly away from \(E_{\rm res}\).

We have also checked the contribution from the $t$-channel process shown in Eq.~\eqref{t-channel}.
As shown in Fig.~\ref{fig:Charged_Higgs_Glashow_cross}, 
 $t$-channel cross section is more than three orders of magnitude smaller than the $s$-channel cross section near the s-channel resonant region.
We therefore neglect this $t$-channel contribution in the following discussion. 

The corresponding cross sections in Model II are shown in the lower panel in Fig.~\ref{fig:Charged_Higgs_Glashow_cross}.
The qualitative behavior is the same as in Model I.
In the resonance region, the $t$-channel cross section is much smaller than the $s$-channel one, so it is also neglected in Model II.
Comparing the upper and lower panels of Figs.~\ref{fig:Charged_Higgs_Glashow_cross}, one finds that the resonance in Model II  is broader and has a lower peak cross section.
The broader resonance is due to the larger total width in Model II, which originates from the additional decay channels in Eq.~\eqref{eq: Charged_Higgs_width_Model-II}.
The relatively lower peak value is caused by the reduced branching ratios for both initial and final states, $\mathrm{Br}(H^-\to e^-\bar{\nu}_e)=\mathrm{Br}(H^-\to\mu^-\bar{\nu}_\mu)=1/9$, as indicated by Eq.~\eqref{eq: Charged_Higgs_cross_section}.

\begin{figure}[t]
    \centering
    \includegraphics[width=0.48\textwidth]{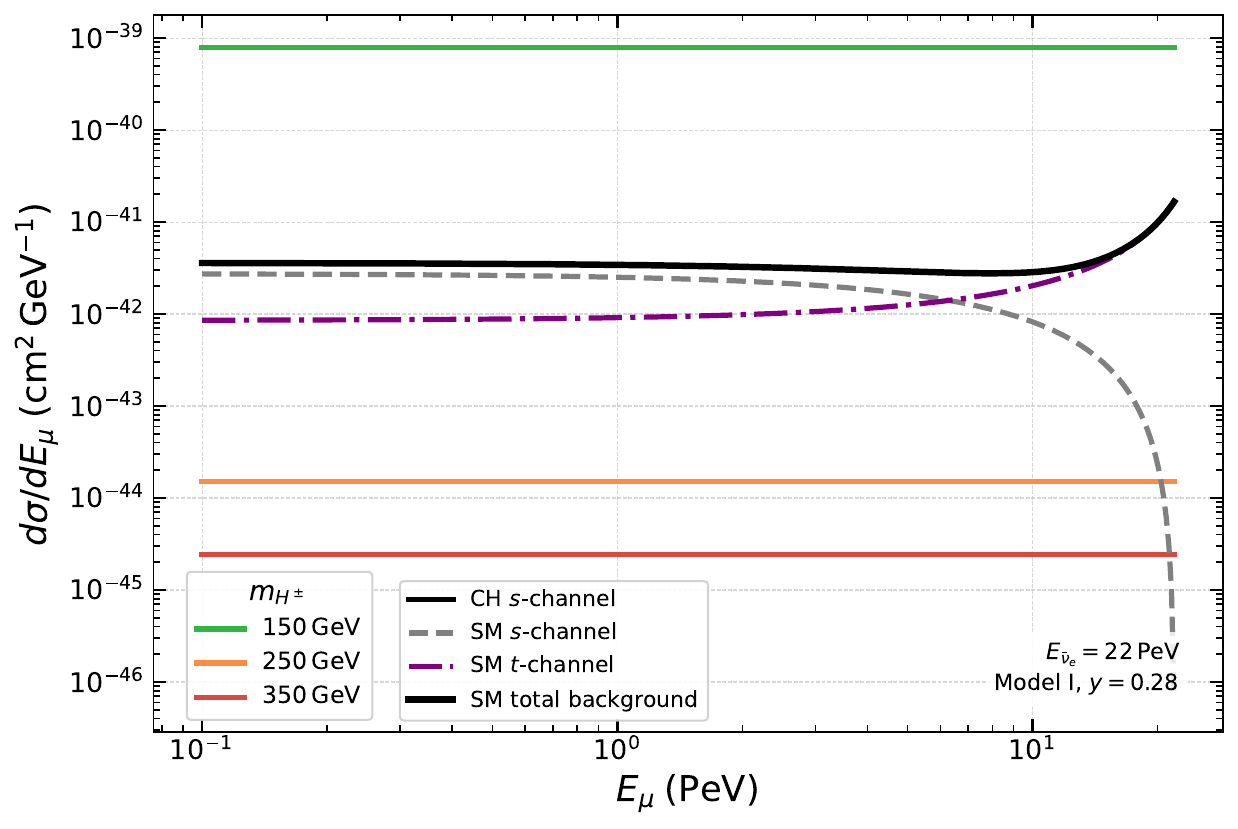}
    \includegraphics[width=0.48\textwidth]
    {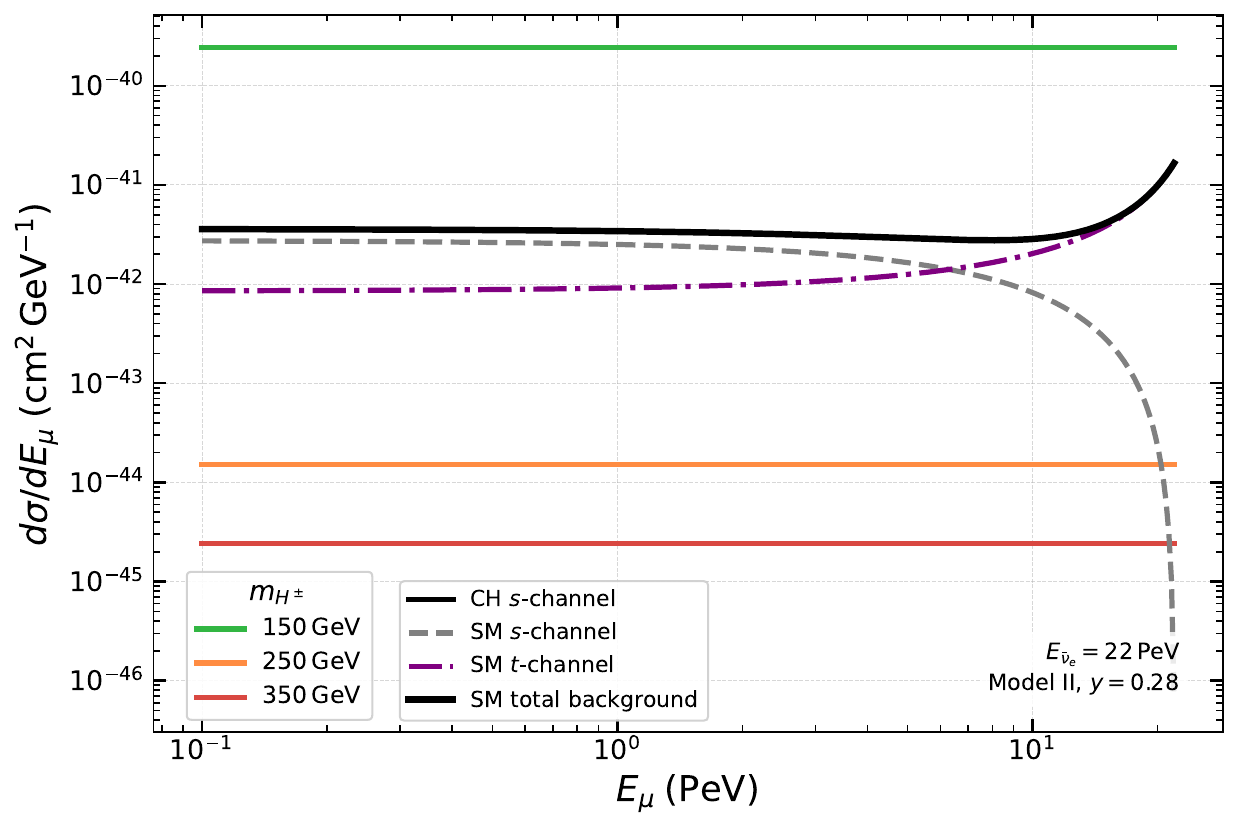}
    \caption{
Differential cross sections as functions of the outgoing muon energy $E_\mu$, evaluated at a fixed incident antineutrino energy $E_{\bar{\nu}_e}=22~{\rm PeV}$, approximately equal to the resonance energy for $m_{H^\pm}=150~{\rm GeV}$.
The green, orange and red solid curves denote the charged Higgs $s$-channel contributions for $m_{H^\pm}=150~\mathrm{GeV}$, $250~\mathrm{GeV}$ and $350~\mathrm{GeV}$, respectively.
The gray dashed and purple dash-dotted curves show the SM $s$- and $t$-channel contributions, respectively, while the black solid curve denotes their sum.
For illustration, we fix the coupling to $y=0.28$.
The upper and lower panels show the results for Model I and Model II, respectively.
}
    \label{fig:dsigmademu}
\end{figure}

To connect the $s$-channel cross sections with the experimentally relevant muon-track signal, we further consider the observable differential cross section with respect to the outgoing muon energy $E_\mu$ in the laboratory frame,
\begin{equation}
\frac{d\sigma}{dE_\mu}
=
\frac{y_e^2y_\mu^2}{32\pi E_{\bar\nu}}\frac{s-m_\mu^2}{\left[(s-m_\mathrm{H}^2)^2+m_\mathrm{H}^2\Gamma_\mathrm{H}^2\right]}.
\label{eq:the_differential_cross_section_with_respect_to_the_outgoing_muon_energy}
\end{equation}
Since this scalar resonance is produced through an $s$-wave annihilation process, the differential cross section is independent of $E_\mu$, yielding a flat muon-energy distribution within the kinematically allowed range at fixed incident antineutrino energy.
For clarity, the differential cross sections for \(m_{H^\pm}=150~\mathrm{GeV}\), \(250~\mathrm{GeV}\) and \(350~\mathrm{GeV}\) are shown in Fig.~\ref{fig:dsigmademu}.
Here, the incident antineutrino energy is fixed to $E_{\bar{\nu}_e}=22~{\rm PeV}$, which approximately corresponds to the resonance energy for $m_{H^\pm}=150~{\rm GeV}$. 
Therefore, the $m_{H^\pm}=150~\mathrm{GeV}$ process is resonantly enhanced, while the $250~\mathrm{GeV}$ and $350~\mathrm{GeV}$ cases are not, leading to differential cross sections that are several orders of magnitude smaller.
Although $d\sigma/dE_\mu$ does not exhibit a characteristic peak, the flux convolved event spectrum develops a characteristic shoulder near $E_\mu\simeq E_{\rm res}$ due to the kinematic upper boundary on $E_\mu$, as discussed in Sec.~\ref{Signature and Sensitivity at IceCube} below.

SM leptonic processes can also produce muon-track events without an accompanying cascade.
The $s$-channel process $\bar{\nu}_e e^- \to W^- \to \mu^- \bar{\nu}_\mu$ is shown by the black solid curve in Fig.~\ref{fig:Charged_Higgs_Glashow_cross}, while the $t$-channel $W$ boson process $\nu_\mu e^-\rightarrow \mu^-\nu_e$ is shown by the gray dotted curve.
At higher incident antineutrino energies, the contribution from the $t$-channel $W$ boson process becomes dominant background, whereas at lower energies the $s$-channel process dominates around the Glashow resonance at $E_\nu\approx 6.3~\mathrm{PeV}$.
The corresponding SM differential cross sections with respect to the outgoing muon energy are shown in Fig.~\ref{fig:dsigmademu}.
Their observable event spectrum will also be discussed in Sec.~\ref{Signature and Sensitivity at IceCube}.

In addition, both the charged Higgs resonance and the SM Glashow resonances can produce $\tau^\pm$, whose decays, $\tau^\pm \to \mu^\pm \nu \bar{\nu}$, also give rise to muon-track events. 
They will be discussed in detail in Sec.~\ref{Signature and Sensitivity at IceCube} and Appendix~\ref{app:tau_decay}.

\subsubsection{Cascade signals}
\label{sec:Cascade signals}
\begin{figure}[t]
    \centering
    \includegraphics[width=0.23\textwidth]{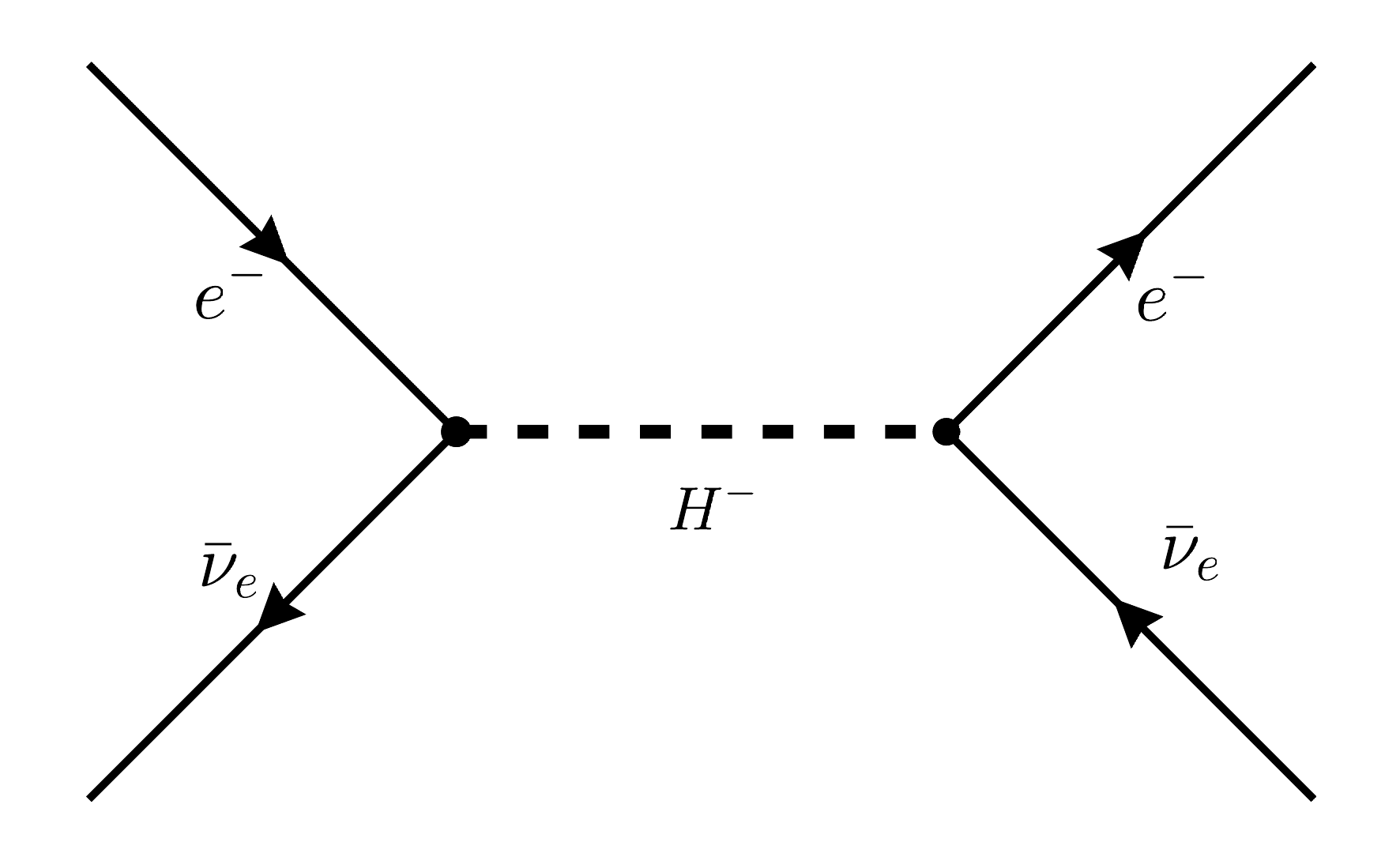}
    \includegraphics[width=0.23\textwidth]{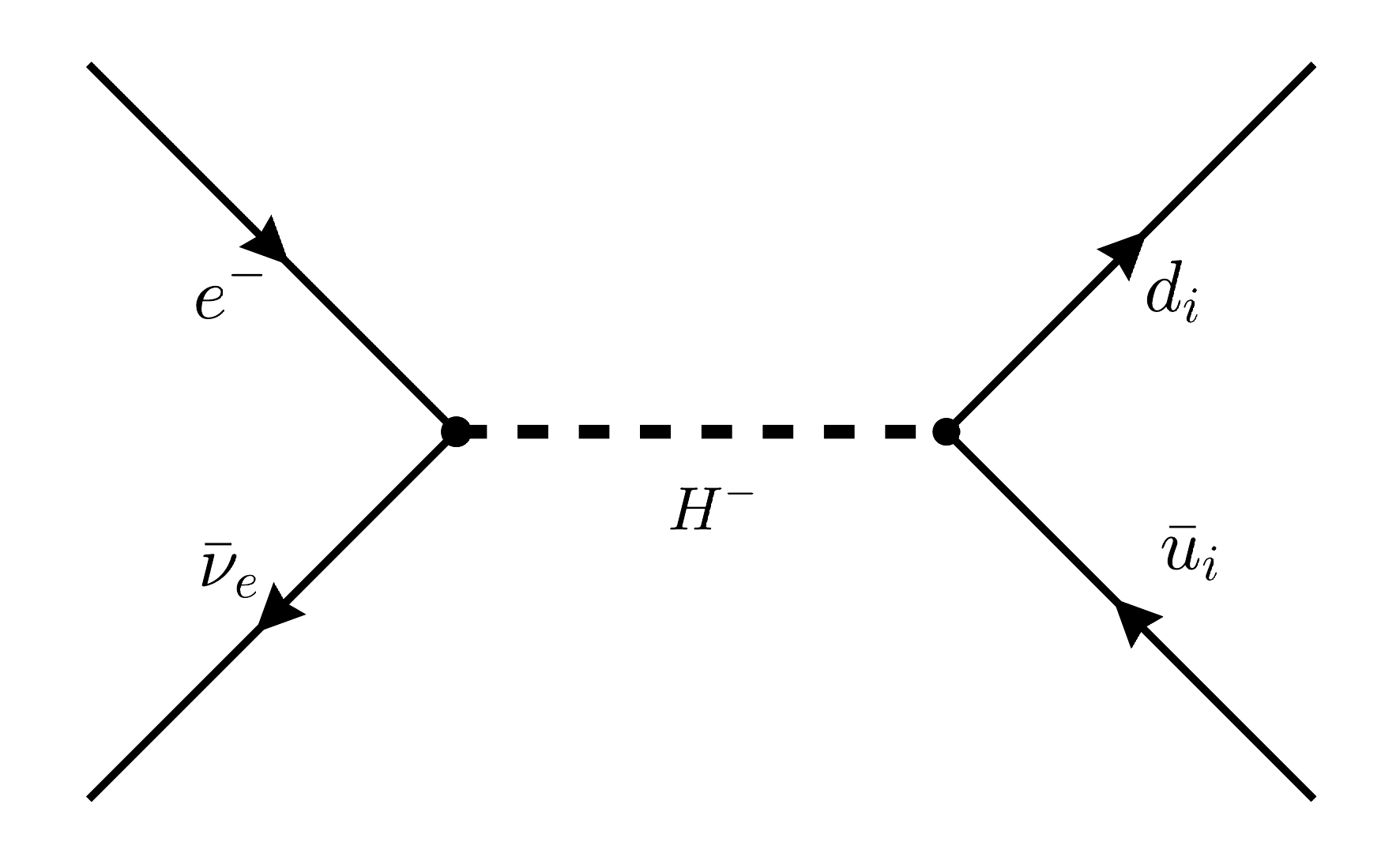}
    \caption{
Feynman diagrams for the charged Higgs processes leading to cascade signals at neutrino telescopes.
The left panel shows $\bar{\nu}_e e^-\to H^-\to e^-\bar{\nu}_e$, which produces an electromagnetic cascade while the right panel shows $\bar{\nu}_e e^-\to H^-\to d_i\bar{u}_i$, which produces hadronic cascade.
In Model II, the additional channels $H^-\to s\bar{c}$ and $H^-\to\tau^-\bar{\nu}_\tau$ are also allowed.
}
    \label{fig:Feynman_cascade}
\end{figure}

\begin{figure}[t]
    \centering
     \includegraphics[width=0.45\textwidth]{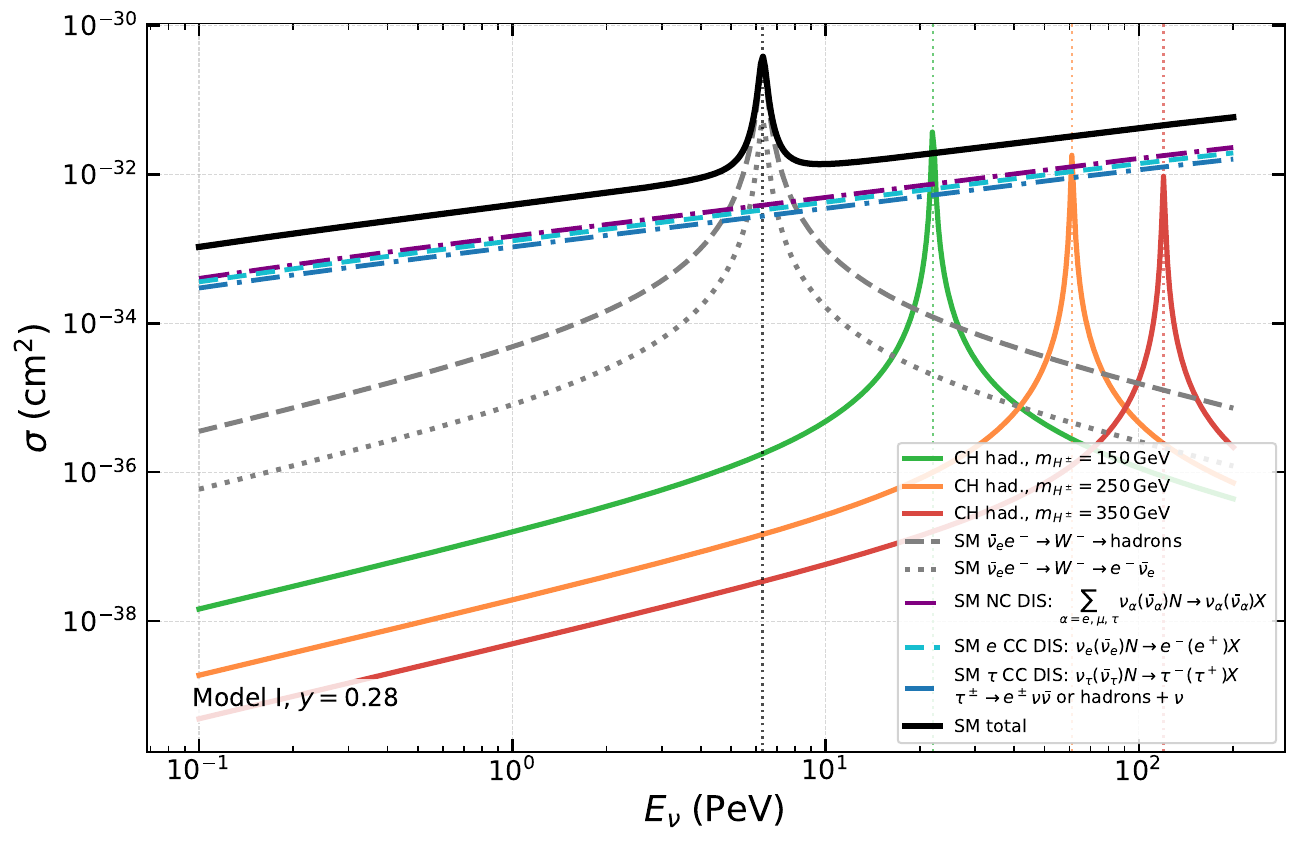}
    \caption{
Total cross sections for the charged Higgs hadronic cascade signal and the relevant SM cascade backgrounds as functions of the incident neutrino energy $E_{\nu}$.
The green, orange, and red solid curves correspond to the charged Higgs hadronic channels with $m_{H^\pm}=150$, $250$, and $350~{\rm GeV}$, respectively.
The gray dashed and dotted curves denotes the SM processes $\bar{\nu}_e e^-\to W^-\to {\rm hadrons}$ and $\bar{\nu}_e e^-\to W^-\to e^-\bar{\nu}_e$, respectively.
The purple, cyan, and blue dash-dotted curves show the SM NC DIS process $\sum_{\alpha=e,\mu,\tau}\nu_\alpha(\bar{\nu}_\alpha)N\to\nu_\alpha(\bar{\nu}_\alpha)X$, electron CC DIS process $\nu_e(\bar{\nu}_e)N\to e^-(e^+)X$, and $\tau$ CC DIS process $\nu_\tau(\bar{\nu}_\tau)N\to\tau^-(\tau^+)X$, respectively.
The black solid curve is the sum of the shown SM cross sections.
The vertical dotted lines indicate the resonance energies of the $W^-$ boson and the charged Higgs bosons.
The results are shown for Model I with $y=0.28$. The cross-section of the t-channel process $\nu_e e^-\to e^- \nu_e$ is much smaller than
the cross section of hadronic cascade events and is not plotted in this figure.}
    \label{fig:sigmaE_nu_model1}
\end{figure}

The charged Higgs can also produce cascade-like final states through the resonant $s$-channel processes
\begin{equation}
\bar{\nu}_e+e^- \to H^- \to e^-+\bar{\nu}_e ,
\label{eq:electronic_cascade}
\end{equation}
and
\begin{equation}
\bar{\nu}_e+e^- \to H^- \to d_i+\bar{u}_i ,
\qquad i=1,2,3,
\label{eq:hadronic_cascade}
\end{equation}
as shown in Fig. \ref{fig:Feynman_cascade}.
The electronic channel produces an electromagnetic cascade, while the quark final states give hadronic cascades.
We combine them as the second signal signature. 
As an example, the cross section of the hadronic cascade process, Eq. (\ref{eq:hadronic_cascade}), as a function of $E_{\nu}$ is shown in Fig.~\ref{fig:sigmaE_nu_model1}.
The cross section of the process (\ref{eq:electronic_cascade}) would have a similar resonance structure. 
Several SM processes can produce similar cascade-like events and therefore contribute to the backgrounds.
The resonant process $\bar{\nu}_e e^-\to W^-\to {\rm hadrons}$ and $\bar{\nu}_e e^-\to W^-\to e^-\bar\nu_e$ provide the background from 
scattering of neutrino on electron targets.
For scattering on nucleon targets, we include the NC DIS processes $\nu_\alpha N\to \nu_\alpha X$ and $
\bar{\nu}_\alpha N\to \bar{\nu}_\alpha X$ as well as the electron and $\tau$ flavor CC DIS processes.
The corresponding cross sections are also shown in Fig.~\ref{fig:sigmaE_nu_model1}.
The cross-section of the SM t-channel process $\nu_e e^-\to e^- \nu_e$, which is not shown here, is of the same order of magnitude of
the cross section of the SM process $\nu_\mu e^-\to \mu^- \nu_e$\cite{Gandhi:1995tf} shown in Fig. \ref{fig:Charged_Higgs_Glashow_cross}
and is much smaller than the cross section of the hadronic cascade events shown in  Fig.~\ref{fig:sigmaE_nu_model1}. 
Similarly,  by inspecting the cross section of the t-channel charged Higgs process $\nu_\mu e^-\to \mu^- \nu_e$ in Fig. \ref{fig:Charged_Higgs_Glashow_cross},
one can find that the contribution of the charged Higgs to the t-channel process $\nu_e e^-\to \nu_e e^- $ is also negligible.

One can find in Fig.~\ref{fig:sigmaE_nu_model1} that,
for the benchmark masses considered, the charged Higgs cross section is below the total SM background over nearly the entire energy range.
Nevertheless, as discussed in Sec.~\ref{Signature and Sensitivity at IceCube}, a statistical analysis of the event excess can still provide some sensitivity to the charged Higgs signal.

We denote $E_{\rm dep}$ as the observable energy deposited in the detector.
For the hadronic final states, we assume that $E_{\rm dep}\simeq E_\nu$.
Thus, the cross section as a function of $E_{\nu}$, shown in Fig.~\ref{fig:sigmaE_nu_model1}, can be mapped onto $\sigma(E_{\rm dep})$, which exhibits the same resonance structure.
This resonant feature provides a distinctive signature for the charged Higgs at neutrino telescopes.
For the electronic channel, since the final state antineutrino is invisible, the deposited energy is entirely contributed by the electromagnetic cascade induced by the outgoing electron.
For simplicity, we assume $E_{\rm dep}\simeq E_e$.
Similar to the discussion for the muon-track channel, the differential cross section of the electronic cascade channel, $d\sigma/dE_{\rm dep}\simeq d\sigma/dE_e$, is also a constant within the kinematically allowed range for a fixed incident neutrino energy.

\subsection{\label{Astrophysical Neutrino Flux Models}Astrophysical Neutrino Flux Models}
The expected event rates, and hence the sensitivity, depend directly on the incident UHE neutrino flux.
Different flux models lead to different predictions.
However, the flux is not directly measurable, but is inferred from observed event samples through reconstruction.
For simplicity, we consider four benchmark flux models, covering different possible flux behaviors.
They are defined as follows.

\begin{figure}[t]
    \centering
    \includegraphics[width=\columnwidth]{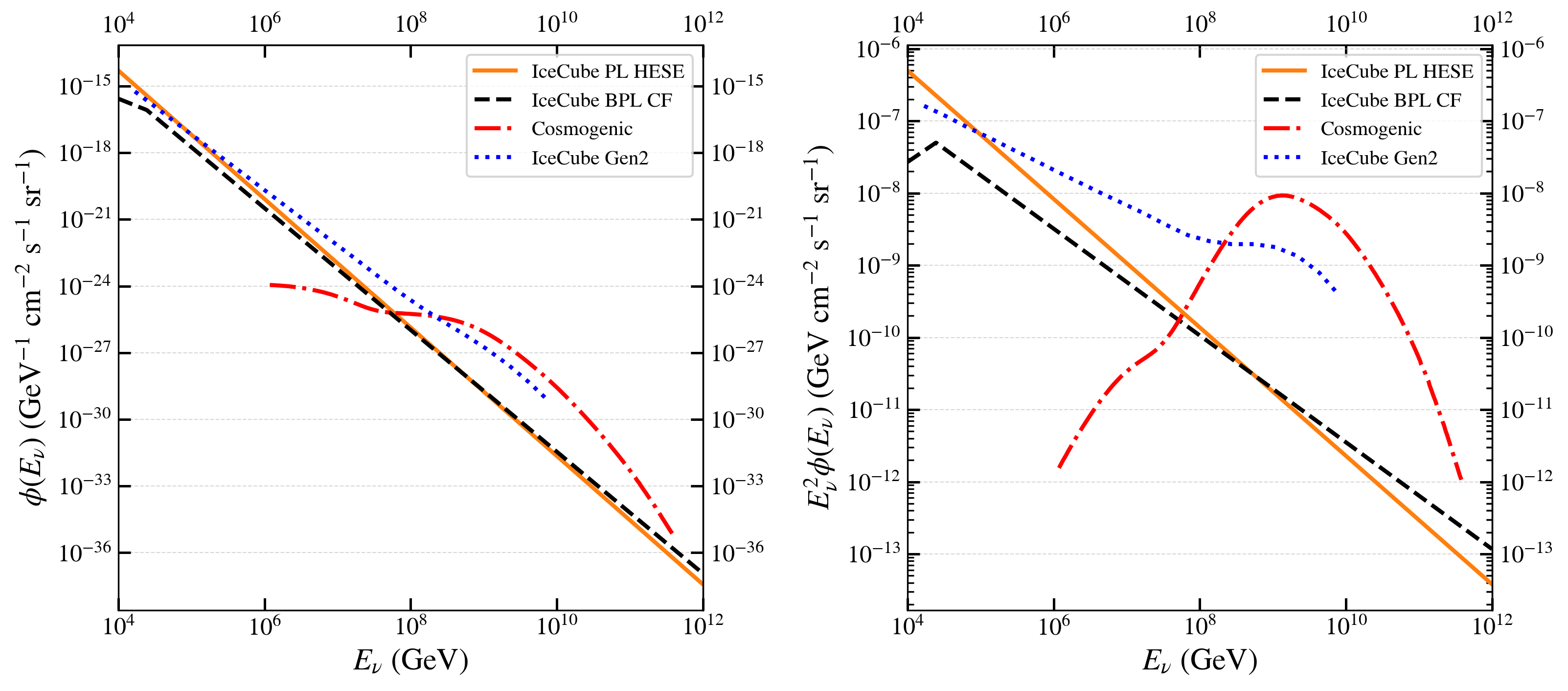}
    \caption{
    All-flavor neutrino flux models used in this work.
    The left panel shows $\Phi_{\nu+\bar{\nu}}(E_\nu)$, while the right panel shows $E_\nu^{2}\Phi_{\nu+\bar{\nu}}(E_\nu)$.
    The \textbf{IceCube PL HESE}, \textbf{IceCube BPL CF}, \textbf{Cosmogenic}, and \textbf{IceCube-Gen2} models are shown by the yellow solid, black dashed, red dash-dotted, and black dotted curves, respectively.
    }
    \label{fig:neutrino_flux}
\end{figure}

The IceCube Collaboration has measured the astrophysical diffuse neutrino flux
using the High-Energy Starting Event (HESE) sample at TeV--PeV energies.
As a benchmark flux model, we adopt their best-fit single-power-law spectrum from Ref.~\cite{Schneider:2019ayi}, and extrapolate it to higher energies relevant for the charged-Higgs resonance.
We denote this UHE neutrino flux model as \textbf{IceCube PL HESE},  shown as the yellow curve in Fig.~\ref{fig:neutrino_flux},
\begin{equation}
\begin{aligned}
\Phi_{\nu+\bar{\nu}}(E_\nu)
&=
6.45\times 10^{-18}
\left(\frac{E_\nu}{100~\mathrm{TeV}}\right)^{-2.89}
\\
&\quad\quad\quad\quad\quad\quad
\left[\mathrm{GeV^{-1}\,cm^{-2}\,s^{-1}\,sr^{-1}}\right].
\end{aligned}
\label{flux: IceCube_PL_HESE}
\end{equation}
Here $\Phi_{\nu+\bar{\nu}}$ denotes the all-flavor diffuse flux at Earth.
Here and hereafter, $E_\nu$ denotes the energy of the incoming neutrino or antineutrino.

In Ref.~\cite{abbasi2025}, a combined fit (hereafter CF) of track and cascade samples was performed within several astrophysical flux model frameworks.
The broken power law model (hereafter BPL) gives the best fit to the data.
We therefore adopt their best-fit broken-power-law spectrum, Eq.~\eqref{flux:IceCube_BPL_CF}, and also extrapolate it to the higher energies relevant for the charged-Higgs resonance.
This defines our second benchmark flux model, denoted as \textbf{IceCube BPL CF}, shown as the black curve in Fig.~\ref{fig:neutrino_flux},
\begin{equation}
\begin{aligned}
\Phi_{\nu+\bar{\nu}}(E_\nu)
={}&
1.77\times10^{-18}
\left(
\frac{10^{4.39}\,\mathrm{GeV}}
{100~\mathrm{TeV}}
\right)^{-2.74}
\\
&\times
\begin{cases}
\left(
\dfrac{E_\nu}{10^{4.39}\,\mathrm{GeV}}
\right)^{-2.74},
&
E_\nu>10^{4.39}\,\mathrm{GeV},
\\[5pt]
\left(
\dfrac{E_\nu}{10^{4.39}\,\mathrm{GeV}}
\right)^{-1.31},
&
E_\nu<10^{4.39}\,\mathrm{GeV},
\end{cases}
\\[-2pt]
&\quad
\left[
\mathrm{GeV^{-1}\,cm^{-2}\,s^{-1}\,sr^{-1}}
\right].
\end{aligned}
\label{flux:IceCube_BPL_CF}
\end{equation}

We then include, as our third flux model, a conservative lower-limit cosmogenic neutrino flux, denoted as \textbf{Cosmogenic}, shown as the red curve in Fig.~\ref{fig:neutrino_flux} \cite{Groth:2021bub, Kampert:2016sqd}.
It is derived from the observed ultra-high-energy cosmic-ray (UHECR) spectrum and the observed UHECR mass composition at Earth \cite{Kampert:2016sqd}.
In the energy range relevant for our study, $E_\nu\sim 10^{7}$--$10^{8}\,\mathrm{GeV}$, this cosmogenic flux lies below the other benchmark fluxes and can therefore be regarded as a conservative choice for our signal estimates.

The fourth benchmark UHE neutrino flux model,
the \textbf{IceCube-Gen2}, is obtained from the expected sensitivity of  the diffuse neutrino flux
for ten years of data taking at IceCube-Gen2~\cite{Aartsen_2021}.

The benchmark fluxes shown in Fig.~\ref{fig:neutrino_flux} are given as total fluxes summed over all flavors.
For the charged Higgs resonance considered in this work, only the $\bar{\nu}_e$ component contributes to the initial state. 
The same $\bar{\nu}_e$ component is used for the SM $s$-channel background, while the $\nu_\mu$ component is used for the $t$-channel background.
Thus, for the incident muon neutrino and electron antineutrino fluxes, we take $\Phi_{\bar{\nu}_e}(E_\nu)=\Phi_{\nu_\mu}(E_\nu)=\Phi_{\nu+\bar{\nu}}(E_\nu)/6$ for all models in Fig.~\ref{fig:neutrino_flux}, assuming equal flavor ratios at Earth and $\nu:\bar{\nu}=1:1$.

\subsection{\label{Signature and Sensitivity at IceCube}Event Rate and Sensitivity at Neutrino Telescope}

Given the benchmark flux models $\Phi_{\bar{\nu}_e}(E_{\nu})$, we calculate the expected event rates with respect to the observable energy $E_{\rm obs} $, namely the outgoing muon energy $E_\mu$ for muon-track signals and the deposited energy $E_{\rm dep}$ for cascade signals.
For a given final state $f$, the differential event rate is
\begin{equation}
\frac{dN_f}{dE_{\rm obs}}
=
N_e T_0
\int_0^{4\pi} d\Omega
\int dE_\nu\,
\frac{d\sigma_f(E_\nu,E_{\rm obs})}{dE_{\rm obs}}\,
\Phi_{\bar{\nu}_e}(E_\nu),
\label{eq:differential_event_rate}
\end{equation}
where $T_0$ is the exposure time, taken to be $T_0 = 3650$ days, corresponding to 10 years of data taking at the neutrino telescope. 
We integrate over $\Omega=4\pi$, assuming full-sky coverage, and neglect detector acceptance and Earth attenuation for simplicity.
The incident $\bar{\nu}_e$ interacts with electrons in neutrino telescope, so the number of targets is $N_e = n_e V_\mathrm{det}$, 
where $n_e$ is the electron number density in ice or in water and $V_\mathrm{det}$ is the detector volume. 
For the SM DIS background with nucleon targets, $N_e$ should be replaced by the nucleon number density.
In the sensitivity estimation below, we consider volumes of $100~{\rm km}^3$, $1000~{\rm km}^3$, $3000~{\rm km}^3$, and $5000~{\rm km}^3$.

\begin{figure}[t]
    \centering
    \includegraphics[width=0.48\textwidth]{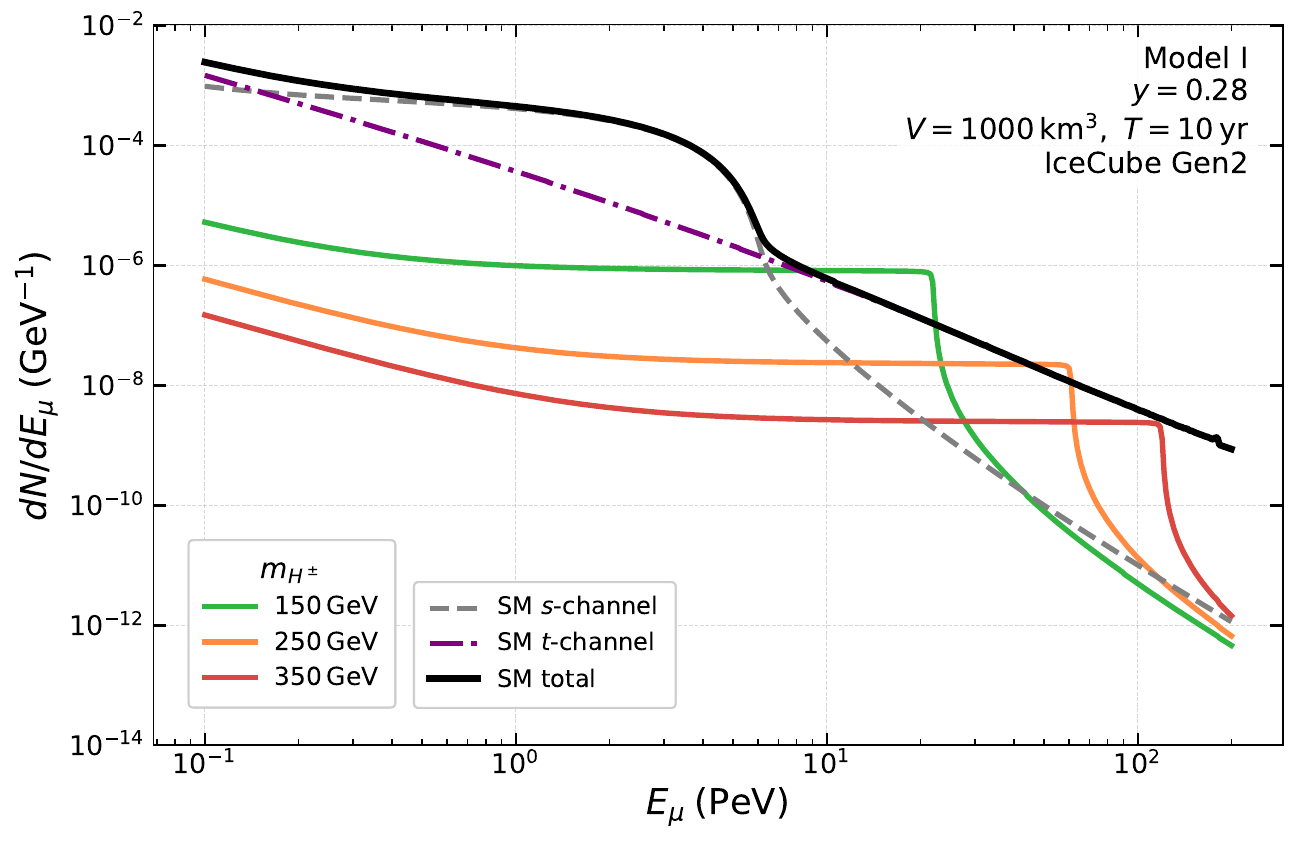}
    \includegraphics[width=0.48\textwidth]{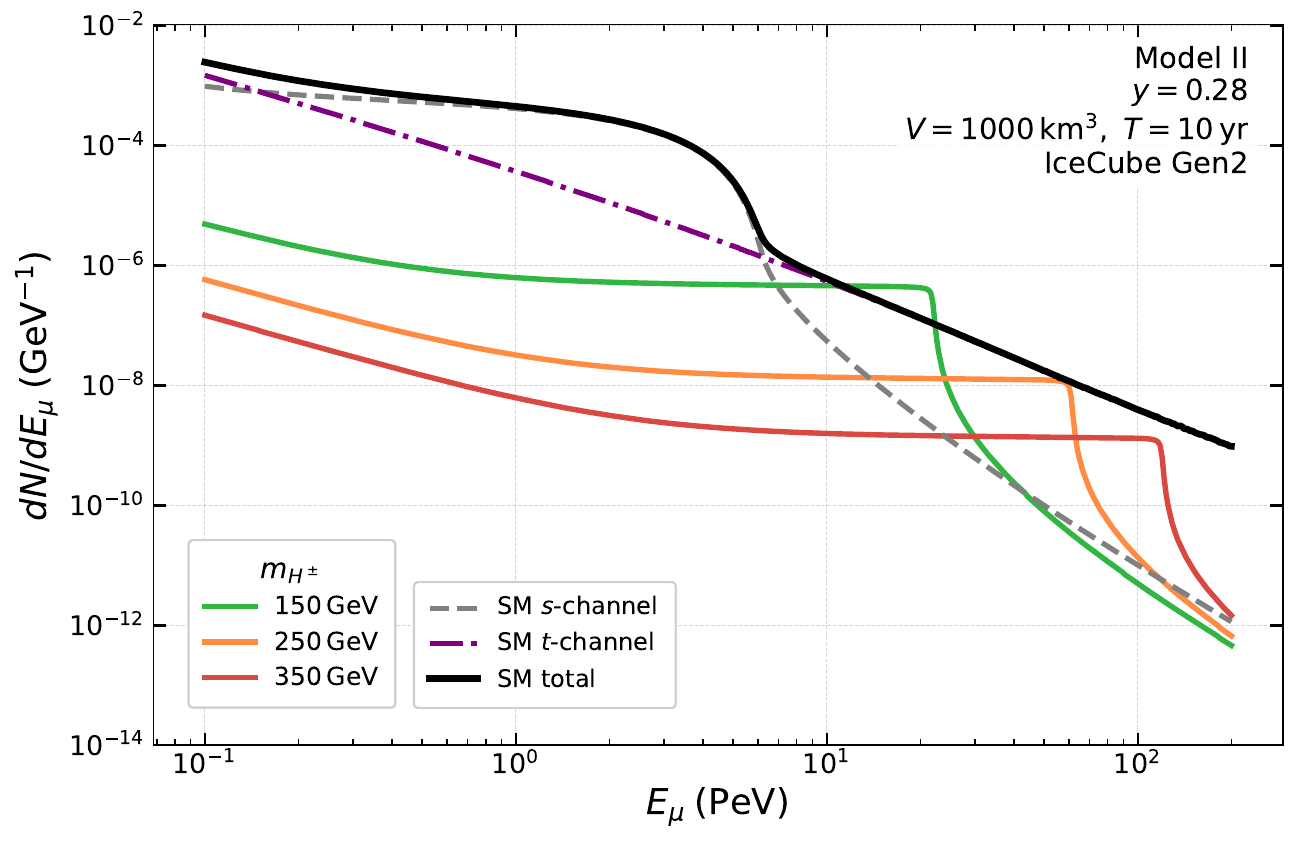}
    \caption{
    Differential muon-track event rates, $dN/dE_\mu$, as functions of the outgoing muon energy $E_\mu$.
    The green, orange, and red solid curves show the charged Higgs $s$-channel signals for $m_{H^\pm}=150$, $250$, and $350~\mathrm{GeV}$, respectively.
    The gray dashed and purple dash-dotted curves denote the SM backgrounds from $\bar{\nu}_e e^- \to W^- \to \mu^- \bar{\nu}_\mu$ and $\nu_\mu e^- \to \mu^- \nu_e$, respectively, while the black solid curve shows their sum.  
    We take the IceCube-Gen2 benchmark flux, $y=0.28$, $V=1000~\mathrm{km}^3$, and an exposure time of $10$ years.
    The upper and lower panels correspond to Model I and Model II, respectively.
    }
    \label{fig:event_rate}
\end{figure}

\subsubsection{Event Rate}
\label{Event Rate}

We first consider the muon-track channel.
The differential event rate as a function of the outgoing muon energy, $dN/dE_\mu$, is shown in Fig.~\ref{fig:event_rate}.
For the charged Higgs signals, the event rates are nearly flat at low $E_\mu$ and develop shoulders at larger $E_\mu$.
This behavior can be understood from the resonant enhancement of the charged Higgs cross section around $E_{\rm res}$ as illustrated in Fig.~\ref{fig:Charged_Higgs_Glashow_cross}.
For $E_\mu < E_{\rm res}$, the scattering process happened at resonant region can contribute to the event rate.
Moreover, at fixed $E_\nu$, the differential cross section $d\sigma/dE_\mu$ is approximately flat, as shown in Fig.~\ref{fig:dsigmademu}.
The resulting event spectrum is therefore nearly flat at low $E_\mu$.
At much lower $E_\mu\ll E_{\rm res}$, however, the event rate is not perfectly flat, because the assumed incident neutrino flux is larger at lower $E_\nu$ and gives non-negligible contributions from the off-resonance scattering process.
For $E_\mu\gtrsim E_{\rm res}$, the incident neutrino must satisfy $E_\nu\gtrsim E_\mu\gtrsim E_{\rm res}$.
The s-channel process shown in Eq.(\ref{eq:s-channel}) then becomes off-resonance and quickly drops down as $E_\mu$ or $E_\nu$ increases.
This suppresses the event rate and produces the shoulder structure.
As $E_{\rm res}\simeq m_{H^\pm}^2/(2m_e)$, the shoulder appears at larger $E_\mu$ for a heavier charged Higgs.
Among the benchmark cases, the shoulder appears at the largest $E_\mu$ for $m_{H^\pm}=350~\mathrm{GeV}$, as shown in Fig.~\ref{fig:event_rate}.
This structure is distinct from the smoothly falling SM background, making the muon-track channel an effective probe of the charged Higgs.
However, for large $m_{H^\pm}$, the differential cross section is suppressed by $m_{H^\pm}^{-4}$, leading to a much lower event rate as shown by the red lines.
Therefore, a heavier charged Higgs is more difficult to probe in the muon track channel.

We have so far considered only the direct production of muons.
In Model II, additional muon-track events can arise from the charged Higgs process ${\bar \nu}_e e^- \to H^-\rightarrow\tau^-\bar{\nu}_\tau$, followed by $\tau^-\rightarrow\mu^-\bar{\nu}_\mu\nu_\tau$.
Similarly, SM processes producing a $\tau$ lepton can contribute to the background.
However, the muon-track contributions from these channels are negligible  in our models compared with the direct contributions and are therefore ignored in our analysis.
Further details are given in Appendix~\ref{app:tau_decay}.

\begin{figure}[t]
    \centering
    \includegraphics[width=0.48\textwidth]{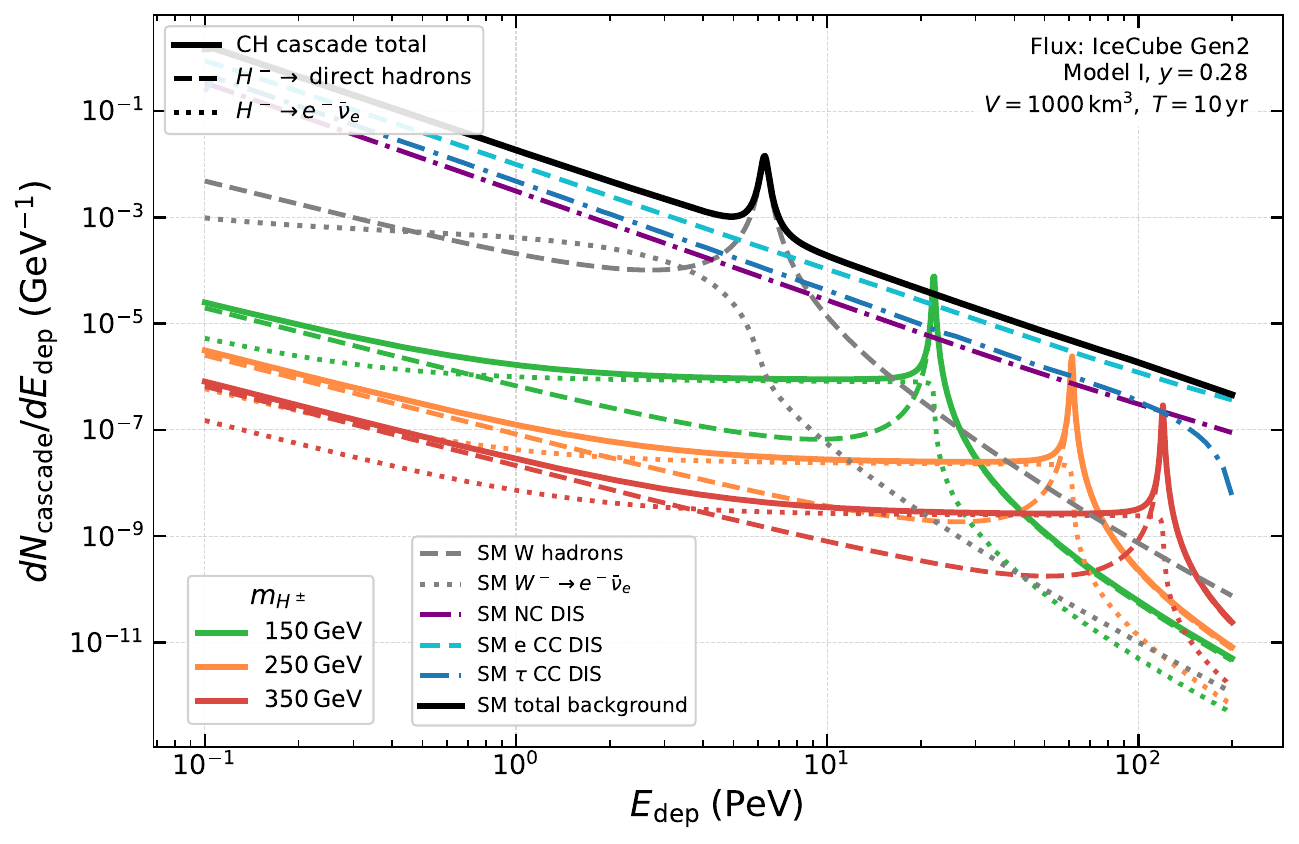}
    \includegraphics[width=0.48\textwidth]{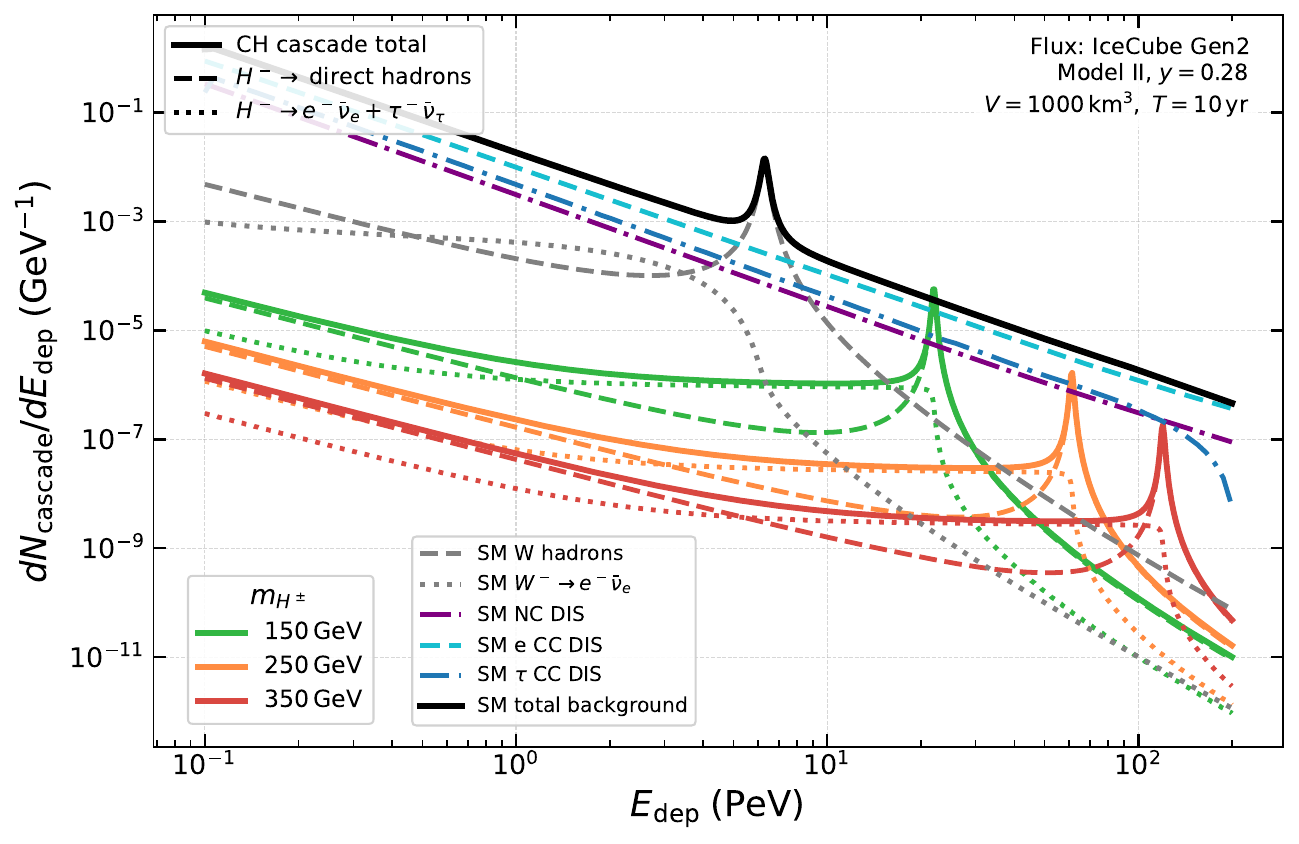}
    \caption{
    Differential rates for cascade events as functions of the energy decomposition $E_{\rm dep}$.
    The green, orange, and red curves represent the charged Higgs signals for $m_{H^\pm}=150$, $250$, and $350~\mathrm{GeV}$.
    For each mass, the dashed, dotted, and solid curves denote the event rates from the charged Higgs hadronic final states, leptonic cascade final states, and their sum, respectively.
    For Model I, the leptonic cascade contribution arises only from $\bar{\nu}_e e^- \to H^- \to e^- \bar{\nu}_e$, whereas for Model II it also includes $\bar{\nu}_e e^- \to H^- \to \tau^- \bar{\nu}_\tau$.
    The gray dashed, gray dotted, purple dash-dotted, cyan dashed, and blue dash-dotted curves denote the same SM background components specified in the caption of Fig.~\ref{fig:sigmaE_nu_model1}, while the black solid curve shows their sum.
    We take the IceCube-Gen2 benchmark flux, $y=0.28$, $V=1000~\mathrm{km}^3$, and an exposure time of $10$ years.
    The upper and lower panels correspond to Model I and Model II, respectively.
    }
    \label{fig:event_rate_cascade}
\end{figure}

As discussed in Sec.~\ref{sec:Cascade signals}, the charged Higgs can also produce cascade events, which also provide a probe of the charged Higgs contribution.
The differential event rates are plotted in Fig. \ref{fig:event_rate_cascade}.
For the hadronic final states, where $E_{\rm dep}\simeq E_\nu$, the charged Higgs cross section retains a resonant dependence on $E_{\rm dep}$.
The corresponding event rate exhibits a narrow peak, which may lead to an excess of cascade events over the SM backgrounds near $E_{\rm dep}\simeq E_{\rm res}$.
Therefore, if a future neutrino telescope observes a significant excess over the SM expectation at an energy well above the $6.3~\mathrm{PeV}$ Glashow resonance, it may be interpreted as a possible resonant charged Higgs signal. 
For a heavier charged Higgs, the resonant peak occurs at a higher energy and has a smaller amplitude as discussed in Sec.~\ref{Muon-track signals}.
This leads to a smaller hadronic cascade event rate and hence a less pronounced excess over the SM background.

Since electromagnetic and hadronic cascades are not readily distinguishable in current neutrino telescopes, 
high-energy electron produced in ${\bar \nu}_e e^- \to H^-\to \bar{\nu}_e e^-$ process
is also treated as a cascade event, for which we take $E_{\rm dep}\simeq E_e$.
$\tau$ can also be produced through the charged Higgs process in Model II.
The electronic and hadronic decay modes of $\tau$, $\tau^-\rightarrow e^-\bar{\nu}_e\nu_\tau$ and $\tau^-\rightarrow{\rm hadrons}+\nu_\tau$,
 also contribute to cascade events, with the corresponding $\tau^+$ modes included for the SM backgrounds.
For illustration, we conservatively approximate $E_{\rm dep}\simeq E_\tau$ for the $\tau$ channel contribution, 
neglecting the missing energy carried away by neutrinos from the $\tau$ decay.
The combined event rates from these leptonic processes are shown by the colored dotted curves in the lower panel in Fig.~\ref{fig:event_rate_cascade}.
It is evident that, around $E_{\rm dep}\simeq E_{\rm res}$, the leptonic cascade contributions are much smaller than the direct hadronic signal.
Away from the resonant region, their event rates lie well below the SM cascade backgrounds.
They are therefore subdominant over the deposited energy range relevant to our analysis, and thus we neglect them in the following discussion.
The validity of the approximation for the $\tau$ channel is examined in detail in Appendix~\ref{app:tau_decay}.
The contribution of the SM charged-current process $\nu_\tau e^-\to\tau^-\nu_e$ is also neglected in the background of cascade events
since it is subdominant as shown in Appendix~\ref{app:tau_decay}.
We have also neglect contribution of the t-channel process $\nu_e e^-\to \nu_e e^-$
because it is negligible as discussed previously for Fig. \ref{fig:sigmaE_nu_model1}.

\subsubsection{Sensitivity}
Having established the signal and SM background event rates, we now estimate the discovery sensitivity of the charged Higgs at future neutrino telescopes.
For each channel, the expected number of events in the $i$-th
energy bin is obtained by integrating the corresponding differential event rate,
\begin{equation}
N_{X,i}
=
\int_{E_i^{\rm min}}^{E_i^{\rm max}}
dE_{\rm rec}\,
\frac{dN_X}{dE_{\rm rec}},
\end{equation}
where $X$ denotes a specific process.
Here, $E_{\rm rec}=E_{\rm dep}$ for cascade events, while $E_{\rm rec}=E_\mu$ for track events.
In the following analysis,
we bin $E_{\rm rec}$ logarithmically from $0.1~{\rm PeV}$ to
$200~{\rm PeV}$, using $N_{\rm bins}$ energy bins.
Then we define the discovery significance as~\cite{Cowan:2010js} 
\begin{equation}
Z_A =
\left\{
2\sum_i
\left[
(N_{s_i}+N_{b_i})\ln\left(1+\frac{N_{s_i}}{N_{b_i}}\right)-N_{s_i}
\right]
\right\}^{1/2} \label{asimov-significance}
\end{equation}
where $N_{s_i}$ and $N_{b_i}$ denote the total expected numbers of charged Higgs signal events and SM background events in each bin, respectively.

For the cascade and track channels, the discovery significances
$Z_A^{\rm cascade}$ and $Z_A^{\rm track}$ are evaluated separately.
Since cascade and muon track events correspond to distinctive event topologies in neutrino telescope, we assume that the two channels are statistically independent.
The combined discovery significance is therefore defined as
\begin{equation}
Z_A^{\rm combined}
=
\sqrt{
\left(Z_A^{\rm cascade}\right)^2+
\left(Z_A^{\rm track}\right)^2
}.
\label{eq:combined_channels}
\end{equation}
For a fixed charged Higgs mass $m_{H^\pm}$, the $n\sigma$ discovery sensitivity in a specific channel $j\in\{{\rm cascade},{\rm track},{\rm combined}\}$ is defined as the value of Yukawa coupling $y$ satisfying
\begin{equation}
Z_{A,j}(m_{H^\pm},y)=n.
\end{equation}
We present both the $3\sigma$ and $5\sigma$ sensitivities in the following. 
We refer to the $5\sigma$ sensitivity as the discovery sensitivity, which constitutes the main result of this work, while the $3\sigma$ sensitivity is shown for comparison.

\begin{figure}[t]
    \centering
    \includegraphics[width=0.48\textwidth]{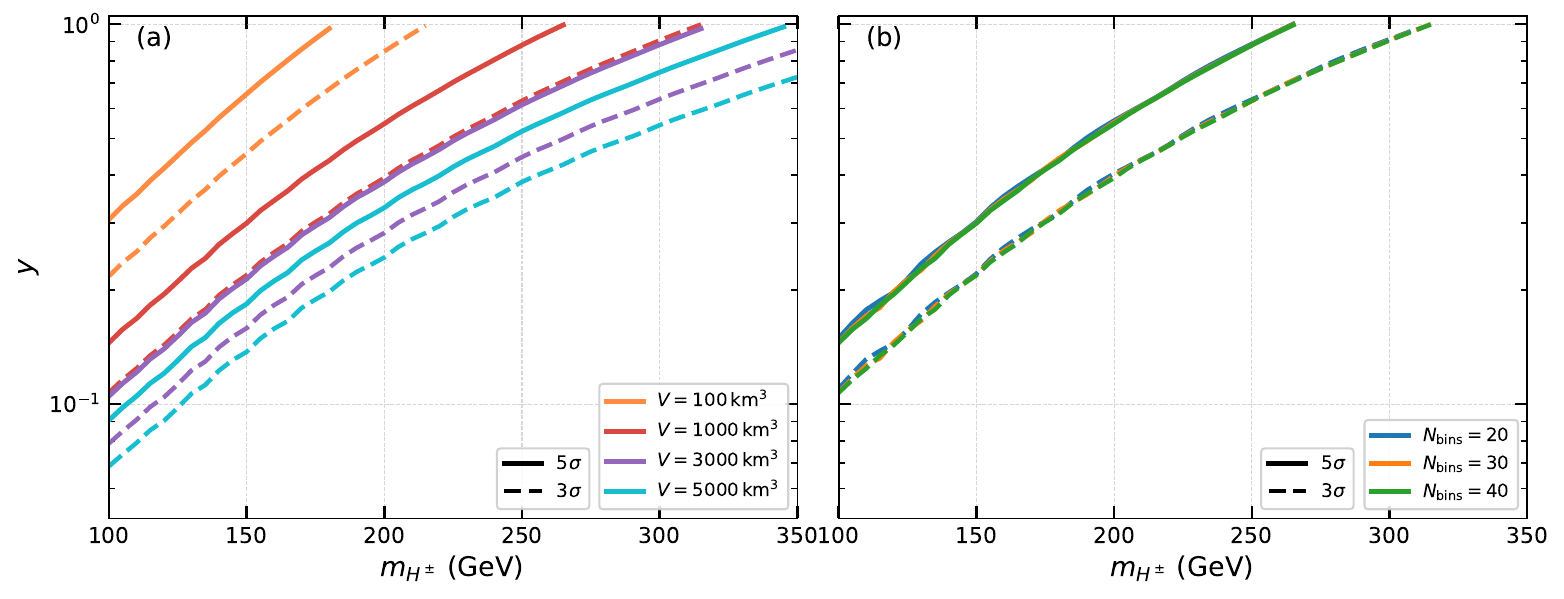}
    \includegraphics[width=0.48\textwidth]{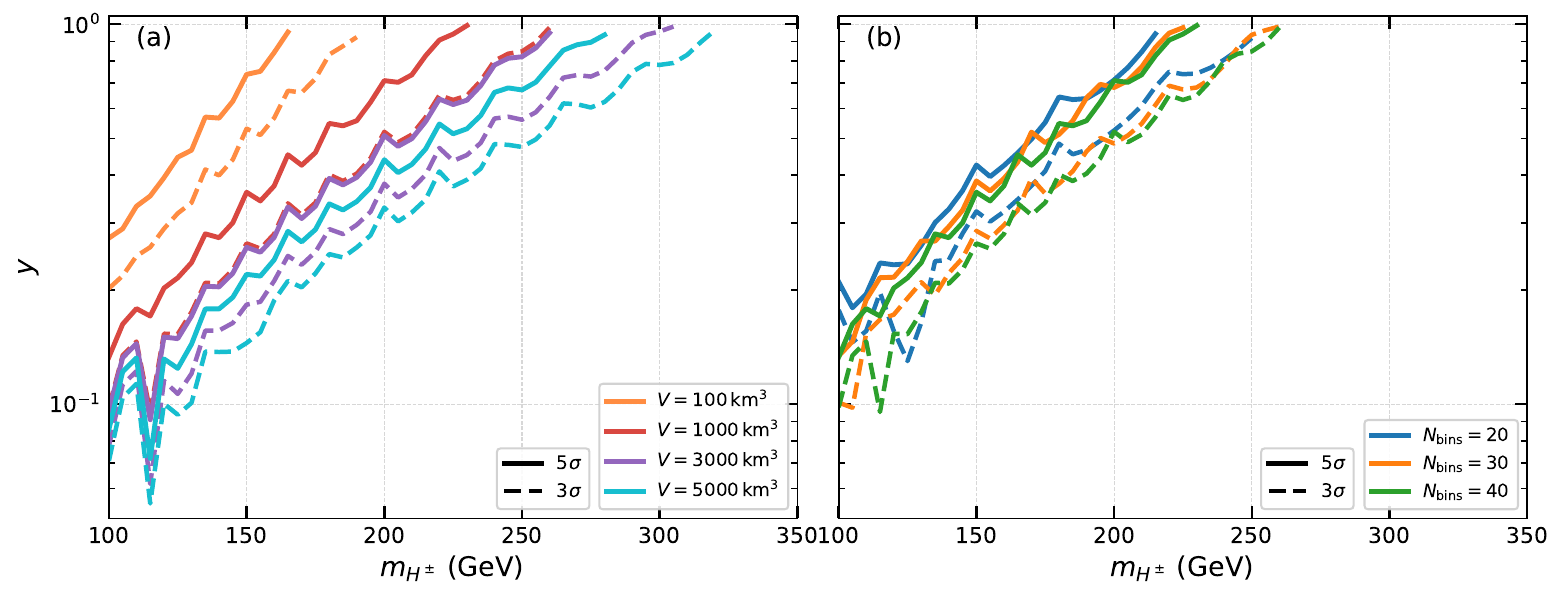}
    \includegraphics[width=0.48\textwidth]{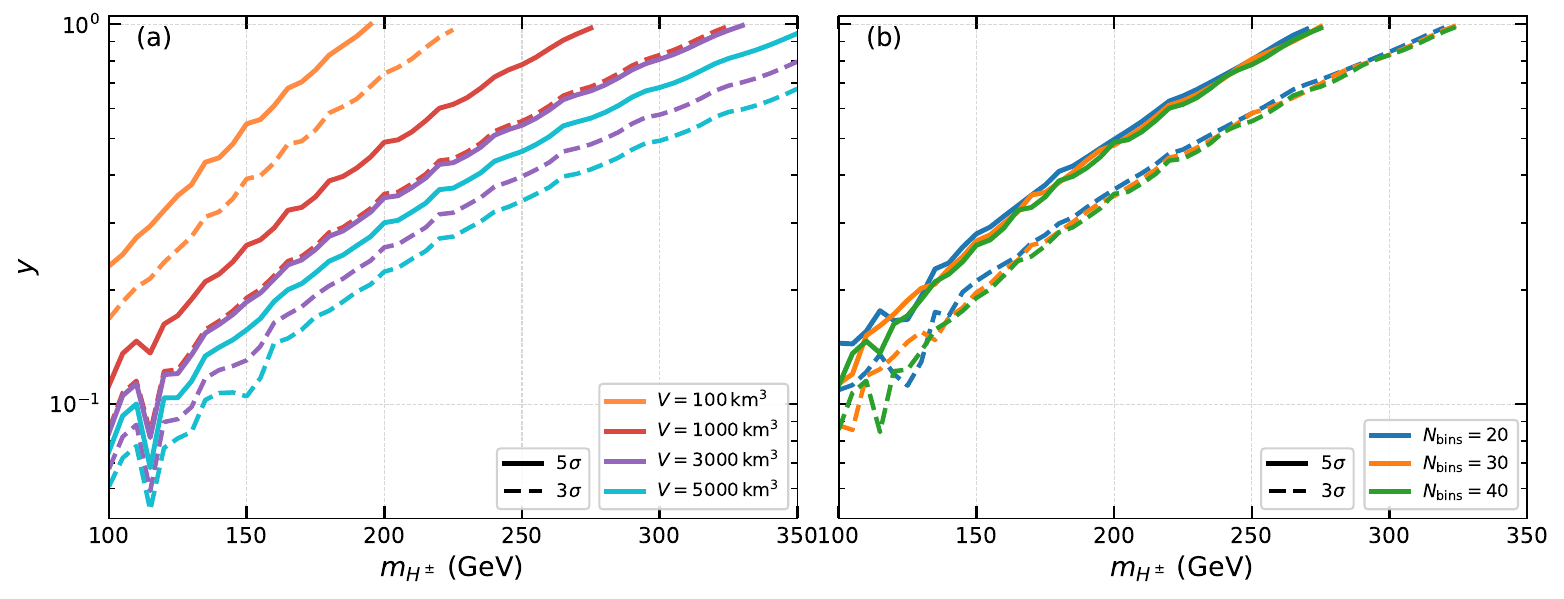}
    \caption{
    Sensitivities as functions of the charged Higgs mass $m_{H^\pm}$.
    Solid and dashed curves show the $5\sigma$ and $3\sigma$ sensitivities, respectively, with an exposure time of $T_0=10$ yr.
    The upper, middle, and lower rows correspond to the muon track, cascade, and combined channels, respectively.
    In the left panels, the sensitivities are shown for different effective volumes with $N_{\rm bin}=40$.
    In the right panels, the dependence on the number of energy bins $N_{\rm bin}$ is shown for a fixed effective volume of $V=1000~{\rm km}^3$.
    All results are obtained for Model I using the IceCube-Gen2 flux.
    }
    \label{fig:channel_sensitivity}
\end{figure}

The $3\sigma$ and $5\sigma$ sensitivities in the muon track channel for Model I, assuming different effective volumes and choices of $N_{\rm bins}$ are shown in the upper panels in Fig.~\ref{fig:channel_sensitivity}.
All sensitivity curves increase with the charged Higgs mass $m_{H^\pm}$.
As discussed in Sec.~\ref{Event Rate}, a heavier charged Higgs has a smaller resonant cross section, leading to a decreased event rate.
Consequently, a larger Yukawa coupling is required to reach the same significance.
The upper left panel shows that increasing the effective volume can substantially improve the sensitivity.
For $V=100~{\rm km}^3$, the $5\sigma$ sensitivity  requires $y=1$ before $m_{H^\pm}$ reaches $200~{\rm GeV}$, 
while for $V=5000~{\rm km}^3$, $5\sigma$ sensitivity can be reached for a charged Higgs boson with $y\simeq1$  and $m_{H^\pm}=350~{\rm GeV}$.
This is because the expected signal and background event numbers in the muon-track channel scale with the effective volume, 
resulting in an improved statistical sensitivity.

The upper right panel in Fig.~\ref{fig:channel_sensitivity} shows that the sensitivity in the muon track channel is
only weakly affected by the choice of $N_{\rm bins}$.
This is because the muon-track event rate is approximately flat over a broad energy
range, and thus does not contain a narrow spectral feature.
This conclusion, however, does not hold for the cascade channel as shown in the middle right panel.
In this channel, a finer binning (or larger $N_{\rm bins}$) can better resolve the resonant peak in the event rate spectrum relative to the smooth SM background, thereby improving the sensitivity.
Moreover, as $m_{H^\pm}$ varies, the resonant peak shifts across the fixed energy bin boundaries.
This leads to artificial wiggles in the sensitivity curves due to finite binning effects.
A comparison of the upper and middle panels shows that the muon track channel provides a stronger sensitivity than the cascade channel.
Therefore, the combined sensitivities shown in the lower panels, obtained using Eq.~\eqref{eq:combined_channels}, are close to those of the muon track channel.

\begin{figure}[t]
    \centering
    \includegraphics[width=0.48\textwidth]{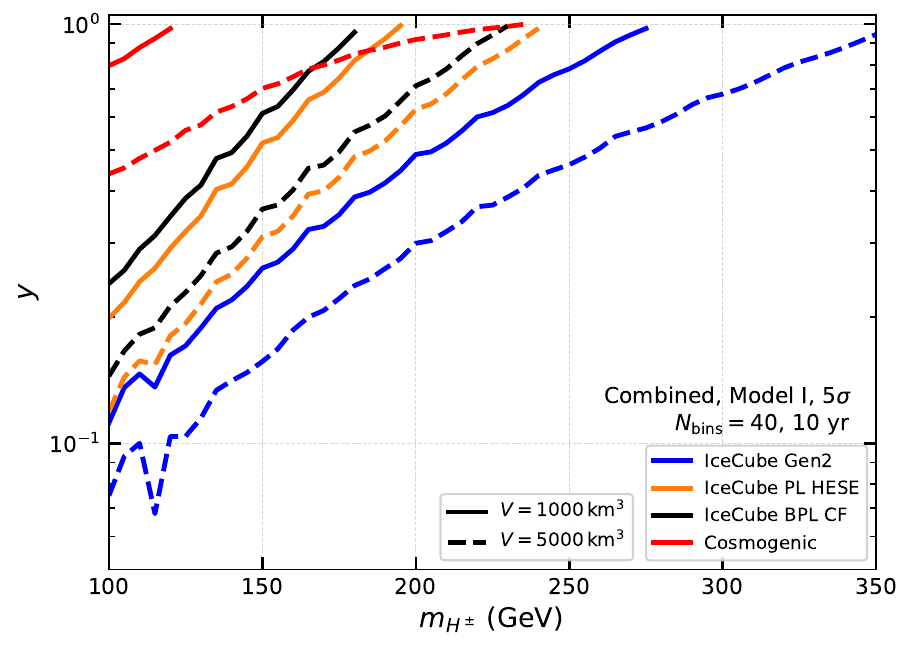}
    \includegraphics[width=0.48\textwidth]{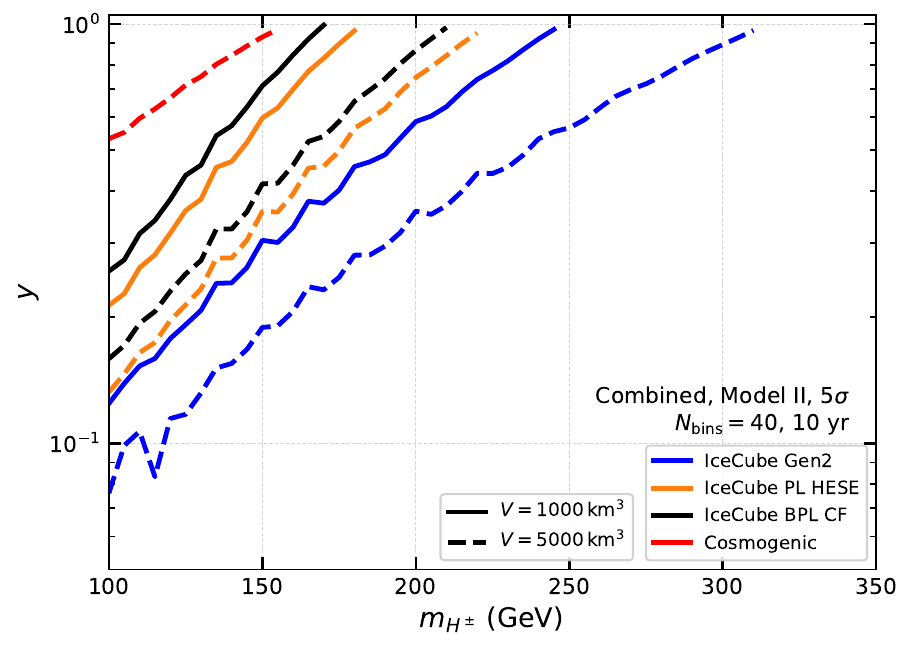}
    \caption{
    $5\sigma$ discovery sensitivities as functions of the charged Higgs $m_{H^\pm}$ for different benchmark neutrino flux models.
    Solid and dashed curves are obtained for effective volumes of $V=1000~{\rm km}^3$ and $5000~{\rm km}^3$, respectively.
    All results assume an exposure time of $T_0 = 10$ yr and $N_{\rm bin}=40$.
    The upper and lower panel correspond to Model I and Model II, respectively.}
    \label{fig:flux_sensitivity}
\end{figure}

Finally, to study the dependence of the sensitivities on the assumed neutrino flux, we repeat the combined analysis for the benchmark flux models introduced in Sec.~\ref{Astrophysical Neutrino Flux Models}.
In addition to Model I considered above, we also include Model II in Fig.~\ref{fig:flux_sensitivity}.
In obtaining these sensitivities, among all the $\tau$-related signal and background contributions, only the $\tau$ CC DIS background is retained and evaluated numerically with \texttt{MadGraph5\underline{~}aMC@NLO}~\cite{Alwall:2014hca} and \texttt{Pythia8}~\cite{Sjostrand:2014zea}.
The other $\tau$-decay contributions from the charged Higgs, Glashow, and the SM $t$-channel processes are neglected.
Further details are provided in Appendix~\ref{app:tau_decay}.

The orange curves for the \textbf{IceCube PL HESE} model are close to the black curves for the \textbf{IceCube BPL CF} model in Fig.~\ref{fig:flux_sensitivity}, since the two flux models are similar.
We therefore focus on the \textbf{IceCube PL HESE} flux in the following discussion.
Moreover, the red dashed curve for the \textbf{Cosmogenic} model crosses the black dashed curve for the \textbf{IceCube BPL CF} model at $m_{H^\pm}\simeq 225~{\rm GeV}$, which corresponds to a resonance energy of approximately $50~{\rm PeV}$.
This behavior originates from the crossing of the two incident neutrino fluxes near this energy as plotted in Fig.~\ref{fig:neutrino_flux}.
For lower $m_{H^\pm}$, the resonance occurs at a lower $E_\nu$, where the \textbf{IceCube BPL CF} flux is larger than the \textbf{Cosmogenic}
flux.It therefore gives a better sensitivity.
For larger $m_{H^\pm}$, the resonance occurs at a higher $E_\nu$, where the \textbf{Cosmogenic} flux becomes larger and gives a better sensitivity.
The above discussion shows that the sensitivity depends strongly on the
incident neutrino flux around the charged Higgs resonance energy.
In a realistic analysis, the neutrino flux should be carefully determined before assessing a possible excess over the SM background.
Such a dedicated analysis is beyond the scope of this work.
A comparison between the upper and lower panels shows that the sensitivities for Model II are substantially weaker than those for
Model I.
In Model I, the charged Higgs resonance has a higher peak, whereas the resonance in Model II is broader because of its larger total width.
Since the energy binning is sufficiently fine to resolve the resonant feature, the higher signal-to-background ratio near the narrower Model I resonance leads to a larger significance.
Consequently,  better sensitivity is achieved in Model I than in Model II.

\section{\label{Collider Signal}Collider Signal}

 Future lepton colliders, such as CEPC~\cite{CEPCStudyGroup:2018ghi} and FCC-ee~\cite{FCC:2018evy}, provide an ideal environment for precision Higgs studies and new physics searches.  In this section, we discuss the production, decay and signals of charged Higgs boson at such facilities.

\subsection{\label{Selection Cuts and Reconstruction Strategy}Selection Cuts and Reconstruction Strategy}

\begin{figure}[t]
    \centering
    \begin{tabular}{cc}
        \centering
        \includegraphics[width=0.22\textwidth]{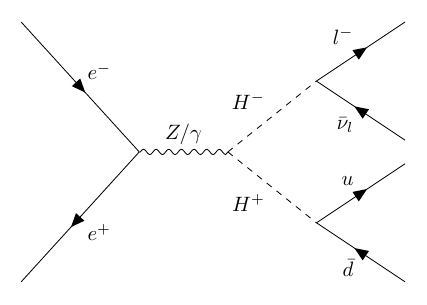} &
         \includegraphics[width=0.22\textwidth]   {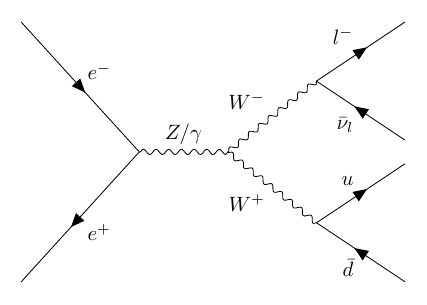} \\
        \includegraphics[width=0.22\textwidth]{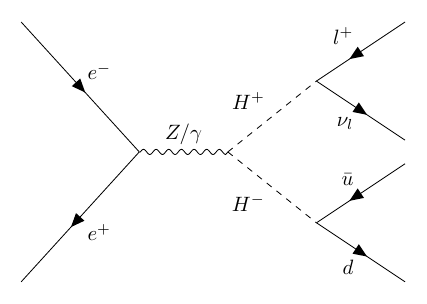} &
        \hfill
        \includegraphics[width=0.22\textwidth]{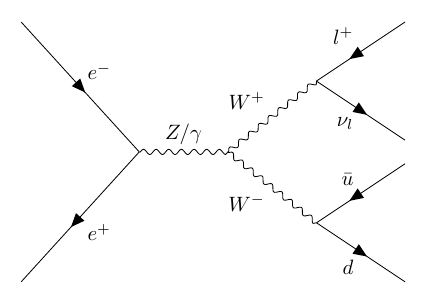} 
    \end{tabular}
    \vspace{0.2cm}
    \caption{
Feynman diagrams of the representative leading-order charged-Higgs pair-production process (left) and the dominant SM background processes (right).
}
    \label{fig_process}
\end{figure}

We focus on the benchmark signal events with two jets, a charged lepton and missing energy at lepton colliders
\begin{equation} \label{eq:colliderprocess}
 e^++ e^- \rightarrow \bar{\nu}_l l^- jj, \, \nu_l l^+ jj
 \end{equation}
which can be mediated by a charged Higgs pair produced  
dominantly through $s-$channel $\gamma/Z$ gauge interaction. 
The most relevant Feynman diagrams for signal and background processes are shown in Fig.~\ref{fig_process}. 
Two invariant masses, $m_{jj}$ and $m_{\bar{\nu}_l l}$,
can be reconstructed in the $\bar{\nu}_l l^- jj$ events using the two jets, and the charged lepton and the missing four momentum.
Similarly, one can also construct two invariant mass for  $\nu_l l^+ jj$ events.
Depending on the charged Higgs mass relative to the center-of-mass energy of the collider, the signal processes exhibit distinct resonant configurations, including double-resonant, single-resonant, and off-shell contributions. 
The corresponding reconstruction strategies and search sensitivities therefore vary accordingly. 
We employ both double-sided and single-sided reconstruction strategies to improve the overall signal acceptance.
The former reconstructs both decay channels using the invariant masses $m_{jj}$ and $m_{\bar{\nu}\ell}$, while the latter retains events with incomplete reconstruction on one of the two channels.

\textbf{Pair On-Shell Region}
In this region, both of the charged Higgs pair are produced on-shell. 
The production rate of charged Higgs pair is expected to be dominated by gauge interaction and independent of the Yukawa couplings of the charged Higgs boson,
as long as the charged Higgs bosons decay inside the detector. 
In this case, both of the reconstructed invariant mass distribution of $m_{jj}$ and of $m_{\bar{\nu}_l l}$  
peak around the mass of the charged Higgs boson. 
In addition, the reconstructed masses from the two reconstructed channels are strongly correlated, 
which motivates the use of the strategy of double-sided reconstruction for both the $jj$ channel and
the ${\bar \nu}_l l^-$ or $\nu_l l^+$ channel to enhance signal-to-noise ratio in analysis shown below. 

\begin{figure}
    \centering
    \includegraphics[width=0.45\linewidth]{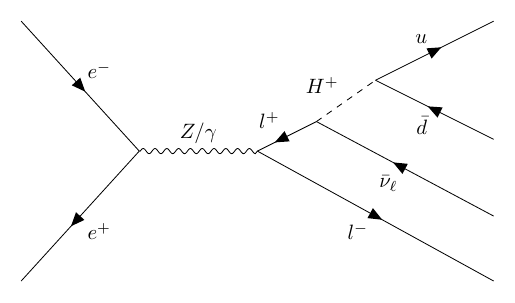}
    \includegraphics[width=0.45\linewidth]{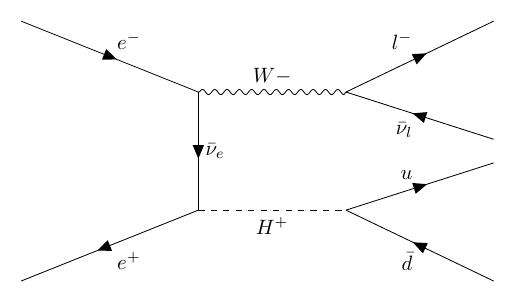}
    \caption{Feynman diagrams of the representative single charged Higgs production process.}
    \label{fig:singleRes}
\end{figure}

\textbf{Single Resonant Region}
As the mass of charged Higgs increases, the center of mass energy of collider is insufficient to ensure both charged Higgs bosons on-shell. 
The signal enters a single-resonant regime, where only one charged Higgs boson can be produced on shell.
The double Higgs production process shown in Fig. \ref{fig_process} is then strongly suppressed because one charged Higgs boson has to be produced off shell.
In this case, the associated single Higgs production process through $W^{\pm}H^{\mp}$ channel, shown e.g. in Fig. \ref{fig:singleRes}, can provide important contributions.
In this case, the double-sided reconstruction won't bring about many benefits comparing with
the single-sided reconstruction.

\textbf{Off-Shell Region}
When the mass of the charged Higgs exceeds the center of mass energy of collider, no on-shell charged Higgs would be produced, and the signal cross section is strongly suppressed.

It is worth emphasizing that charged Higgs pair production probes a qualitatively different aspect of the underlying theory compared to single charged Higgs production.   The latter is highly sensitive to the strength of the Yukawa coupling because 
the single charged Higgs boson is produced through a vertex of Yukawa coupling, as can be seen in Fig.~\ref{fig:singleRes}, whereas in the former case
the charged Higgs boson pair is produced through a vertex of gauge interaction.

To investigate the discovery potential of charged Higgs boson at the future Higgs factories,  the following Monte Carlo (MC) simulations are performed. The initial parton level events including both signal and background events are generated at leading order using \texttt{MadGraph5\underline{~}aMC@NLO}~\cite{Alwall:2014hca}. These events are then processed for hadronization and parton showering using \texttt{Pythia8}~\cite{Sjostrand:2014zea}. The detector effects are simulated using \texttt{Delphes}~\cite{deFavereau:2013fsa}. 

\begin{figure}[!htb]
	\centering 
	\includegraphics[width=0.4\textwidth, angle=0]{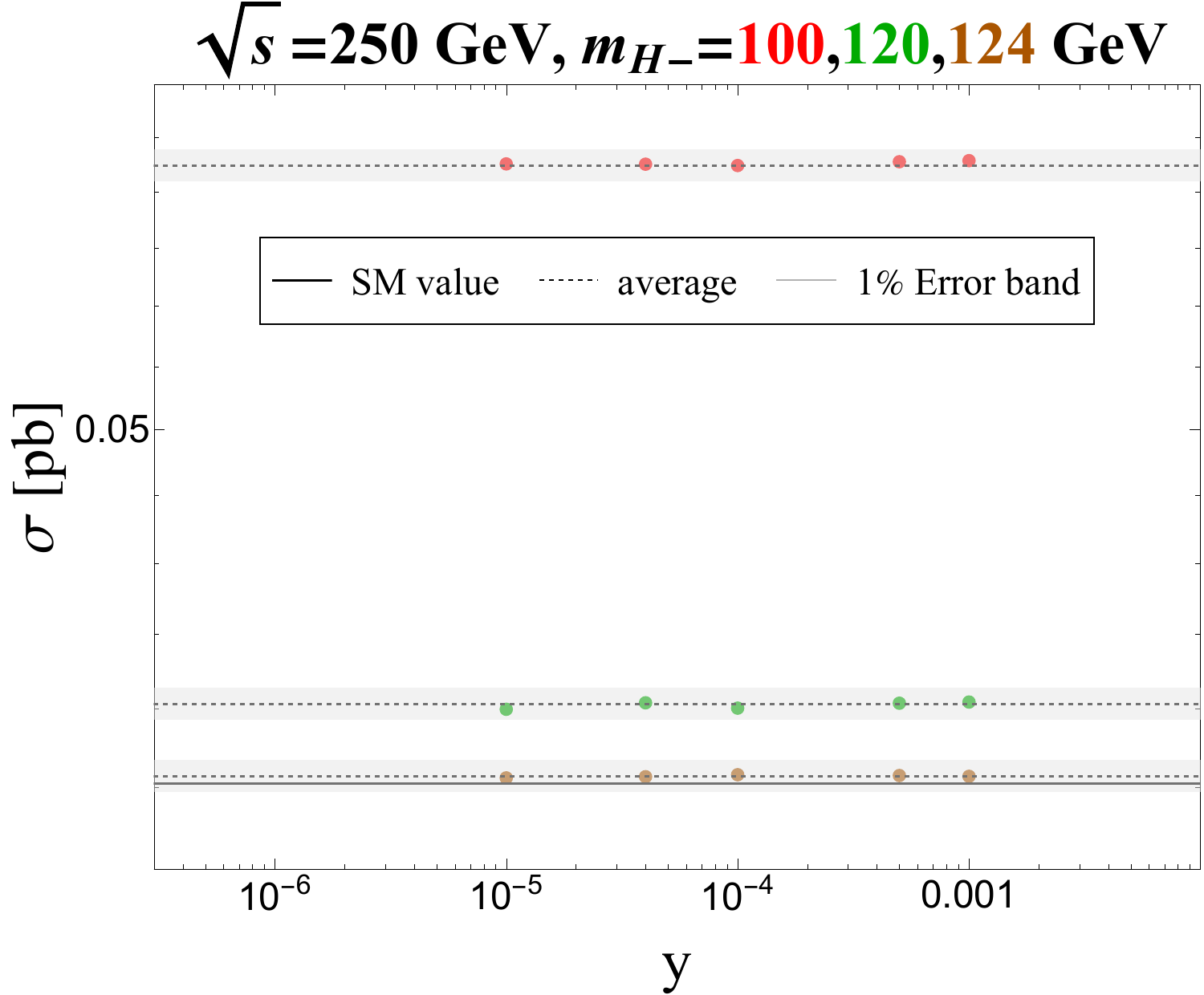}\,
	\caption{ Total cross sections for the 
     $e^+ e^- \rightarrow \bar{\nu}_l l^- \bar{d}u$ processes with $l=e,~\mu$  at $\sqrt{s}=250~{\rm{GeV}}$ after event selection. 
    The black solid line denotes the SM result. Data of the colored dots are obtained from MC simulations scanning over the charged Higgs mass and Yukawa coupling.   The gray band denotes the MC uncertainty which is $1\%$ considered here
    and the dashed line shows the average over the sample values obtained with different Yukawa couplings.  } 
	\label{fig_sig2}%
\end{figure}

For the reconstruction, events are required to contain at least two jets and one charged lepton.
For both $\sqrt{s}=250~{\rm{GeV}}$ and $350~{\rm{GeV}}$ we adopt the following basic cuts for lepton and jets to select the events, 
\begin{align} \label{basicCut1}
	&p_T^j > 20~{\rm{GeV}},~ |\eta^j| < 2.5,~ \Delta R_{jj} > 0.4,\\ \label{basicCut2}
	&p_T^l > 10~{\rm{GeV}},~ |\eta^l| < 2.5,~ \Delta R_{ll} > 0.4,~ E_T^{\rm{miss}} > 10~{\rm{GeV}}.  
\end{align}
In the pair on-shell production regime, as shown in Fig.~\ref{fig_sig2}, the signal rate remains stable as the Yukawa coupling decreases, allowing the charged Higgs mass to be robustly reconstructed using the invariant mass of its visible decay products.
This is because the production rate of charged Higgs pair is dominated by gauge interaction in this case, and as long as the Yukawa coupling of the
charged Higgs is not too small to decay outside the detectors, the cross section of  the $e^++ e^- \rightarrow  \bar{\nu}_l l^- jj$ process
is independent of the Yukawa coupling of the charged Higgs.
 To ensure that the decay happens inside the detector, we impose that the decay length is shorter than the characteristic detector size, taken to be of order 1 m. Charged Higgs bosons with sufficiently small Yukawa couplings to decay outside the detector can be detected through charged tracks.

We consider both double-sided and single-sided reconstruction in the pair on-shell region.
To suppress the background, we require the dijet invariant mass to satisfy 
\begin{align}\label{selCut1}
    90~{\rm{GeV}} < m_{jj} < \sqrt{s}~{\rm{GeV}}
\end{align}
in both reconstruction cases. 
It is used to exclude background events coming from the decay of on-shell $W$ boson in the background processes and improve the signal significance $S/\sqrt{B}$. 
The results are shown in Fig.~\ref{fig_Z}.

In the mass region above half of the center-of-mass energy, i.e., $m_{H^{\pm}} > \sqrt{s}/2$, 
which corresponds to $m_{H^{\pm}} > 125~{\rm{GeV}}$ for $\sqrt{s}=250~{\rm{GeV}}$ and $m_{H^{\pm}} > 175~{\rm{GeV}}$ for $350~{\rm{GeV}}$, the selection cuts are chosen to be \begin{align} \label{selCut2}
   100~{\rm{GeV}} < m_{jj} < \sqrt{s}~{\rm{GeV}}
\end{align}
for consideration of both the suppression of the background and the reduction of the uncertainty caused by MC simulation. The charged Higgs mass is reconstructed using the invariant mass of the visible decay products from jet final state.

The well-used expected discovery significance value is
\begin{align}\label{eq:conven}
	Z = \frac{N_s}{\sqrt{N_b}}, 
\end{align}
where $N_s$ and $N_b$ are the event number of the signal and backgrounds respectively. 
In more general cases,  we can use significance formula, Eq. (\ref{asimov-significance}),  for which Eq. (\ref{eq:conven}) can be obtained from 
Eq. (\ref{asimov-significance}) in the limit $N_s \ll N_b$ for a single bin.
To account for the statistical fluctuations of the Monte Carlo samples and obtain a more stable significance estimate,
we include a systematic uncertainty in the significance estimation.
When systematic uncertainties on the background estimation are taken into account,  the significance formula can be further generalized by introducing nuisance parameters in the likelihood function. In the presence of a relative background uncertainty $\sigma_B$, the resulting significance is reduced, reflecting the loss of sensitivity due to imperfect background knowledge. We use the Cowan significance formula~\cite{Cowan:2010js} 
\begin{widetext}
\begin{equation}
\label{eq:Cowan}
Z_c =
\sqrt{
2\left[
(N_s + N_b)\ln\left(
\frac{(N_s + N_b)(N_b+\sigma_B^2)}
{N_b^2+(N_s + N_b)\sigma_B^2}
\right)
-
\frac{N_b^2}{\sigma_B^2}
\ln\left(
1+\frac{\sigma_B^2 N_s}
{N_b(N_b+\sigma_B^2)}
\right)
\right]
}.
\end{equation}
\end{widetext}
 We set $\sigma_B=0.01$ which would give a quite mild modification to the value of significance at around $5\sigma$.
Eq.~\eqref{eq:Cowan} reduces to \eqref{eq:conven} in the limit $N_S\ll N_B$ with $\sigma_B\rightarrow 0$.
Effect of energy bins can be taken into account appropriately as in Eq. (\ref{asimov-significance}).

\subsection{\label{Constraints from Current Experiments}Constraints from Current Experiments}

Direct searches at past and present colliders have placed stringent constraints on the parameter space of charged Higgs bosons. 
These constraints include the kinematic limits on the charged Higgs mass and bounds on its effective couplings to SM fermions and gauge bosons. Importantly, many of these limits can be interpreted in a largely model-independent manner in terms of the charged Higgs mass and its Yukawa interactions.

At the LEP, charged Higgs bosons were predominantly searched for via pair production in electron–positron collisions, $e^+e^- \rightarrow H^+H^-$, which proceeds through electroweak gauge interactions and is therefore largely independent of the charged Higgs Yukawa couplings. As a consequence, LEP established a robust lower bound on the charged Higgs mass, at the level of $m_{H^{\pm}}\gtrsim 80~{\rm{GeV}}$~\cite{L3:2003jyb,ALEPH:2013htx}, assuming standard decay modes. 
For comparison, this constraint will be included in the plots shown in the next subsection.

At the LHC, light charged Higgs bosons are primarily searched for in top-quark decays, $t\to bH^+$.
For heavier $H^{\pm}$ masses, searches focus on associated production with top and bottom quarks, $pp\to tbH^\pm$, followed by $H^\pm\to\tau^\pm\nu$.
CMS and ATLAS have reported model-independent limits on the corresponding production rate times branching fraction over $80~\mathrm{GeV}\lesssim m_{H^\pm}\lesssim3~\mathrm{TeV}$ \cite{CMS:2019bfg,ATLAS:2024hya}. 
The same production mechanisms have been probed through the fermionic decays $H^\pm\to tb$, $cb$, and $cs$, using multi-$b$-jet or lepton-plus-jets signatures \cite{ATLAS:2021upq,ATLAS:2023bzb,ATLAS:2024oqu}.
Beyond the channels induced by the fermionic couplings of $H^\pm$, 
complementary searches have considered vector-boson fusion $qq\to qqH^\pm$, followed by $H^\pm\to W^\pm Z$~\cite{CMS:2021wlt,ATLAS:2024txt} and Drell--Yan pair production constrained by recasting the corresponding ATLAS and CMS searches for direct stau-pair production~\cite{ATLAS:2019gti,CMS:2022syk}, respectively. 
Searches for bosonic decays, including $H^\pm\to W^\pm h$, $W^\pm A$, and $W^\pm H$, have also been explored \cite{ATLAS:2024rcu,CMS:2019idx,CMS:2022jqc}.
   
CMS and ATLAS searched for $H^{\pm}\rightarrow \tau^{\pm}\nu$ over the charged Higgs mass ranging from 80 GeV to 3 TeV and found no significant excess~\cite{CMS:2019bfg,ATLAS:2024hya}. 
They set 95$\%$ confidence level limits on $\sigma^{\rm{prod}}_{H^{\pm}} \times {\rm{BR}}({H^{\pm}}\rightarrow \tau^{\pm}\nu)$ as a function of the charged Higgs mass without assuming a specific realization of the charged Higgs sector. 
In our models, the charged Higgs does not couple to third-generation quarks. 
Therefore, both decay $t\to bH^+$ and $tbH^{\pm}$ associated production are absent in the charged Higgs mass range of 80–160 GeV and high masses, respectively. 
Consequently, the CMS and ATLAS limits based on these production channels do not directly constrain our model.
In addition, reproducing the limits for charged Higgs mass larger than 160 GeV requires knowledge of the detector acceptance and selection efficiency.
We therefore cannot directly translate them into a bound on the universal Yukawa coupling.
Nevertheless, we retain them as a model-dependent reference by translating the experimental constraint on $\tan\beta$ using 
\begin{align}
    y_{\tau\nu}^{\mathrm{Type-II}}=\sqrt{2}m_\tau\frac{\tan\beta}{v}
\end{align}
with $v$ the electroweak vacuum expectation value.
This result provides only a reference for the possible constraint on the universal Yukawa coupling $y$ defined in Model II. 
 For Model I, the absence of the coupling to $\tau$ lepton removes the constraint from searches based on the $\tau \nu$ final state. 

\subsection{\label{Result}Result}

\begin{figure}[!htb]
	\centering 
	\includegraphics[width=0.22\textwidth, angle=0]{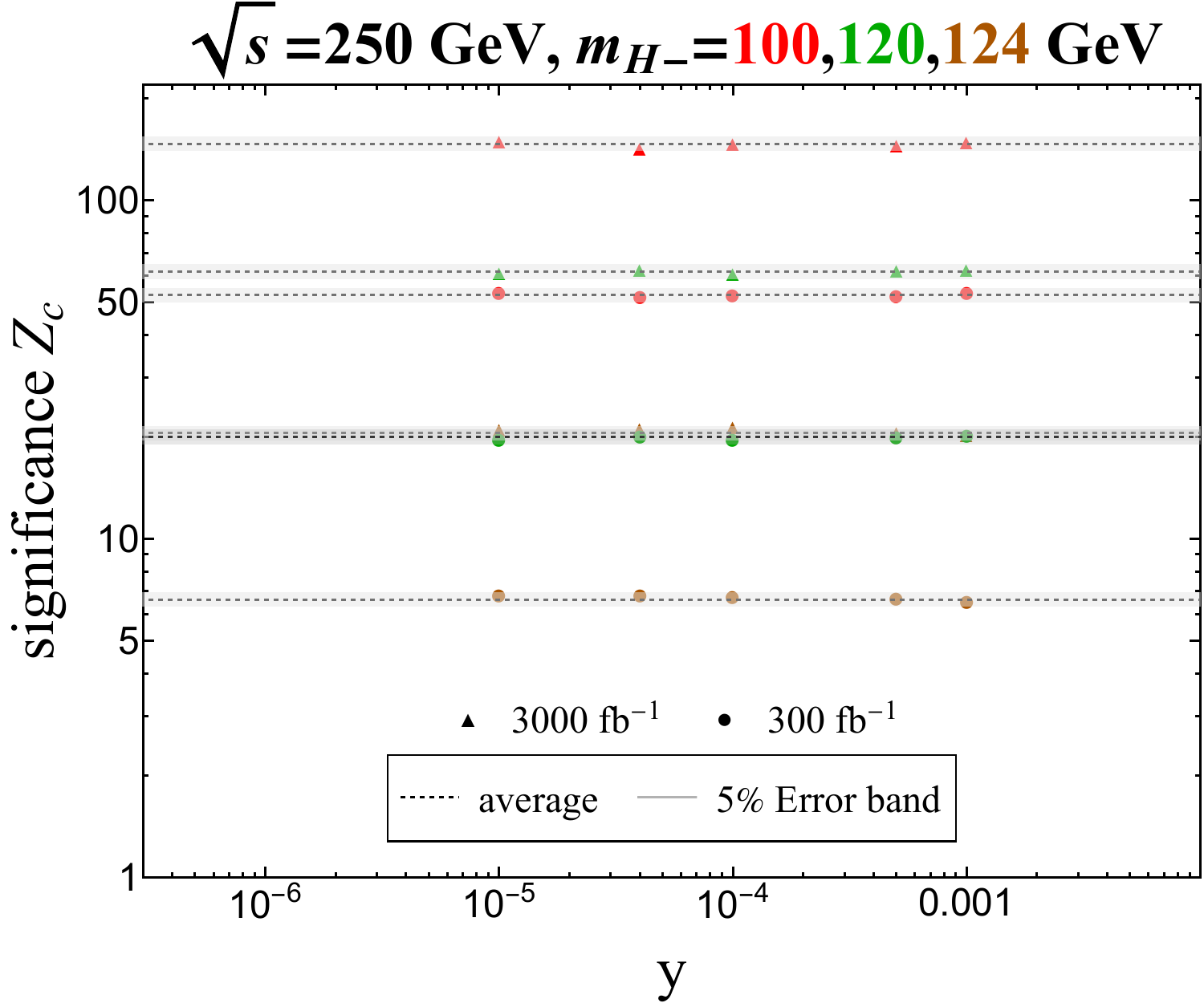}\, \space 
	\includegraphics[width=0.22\textwidth, angle=0]{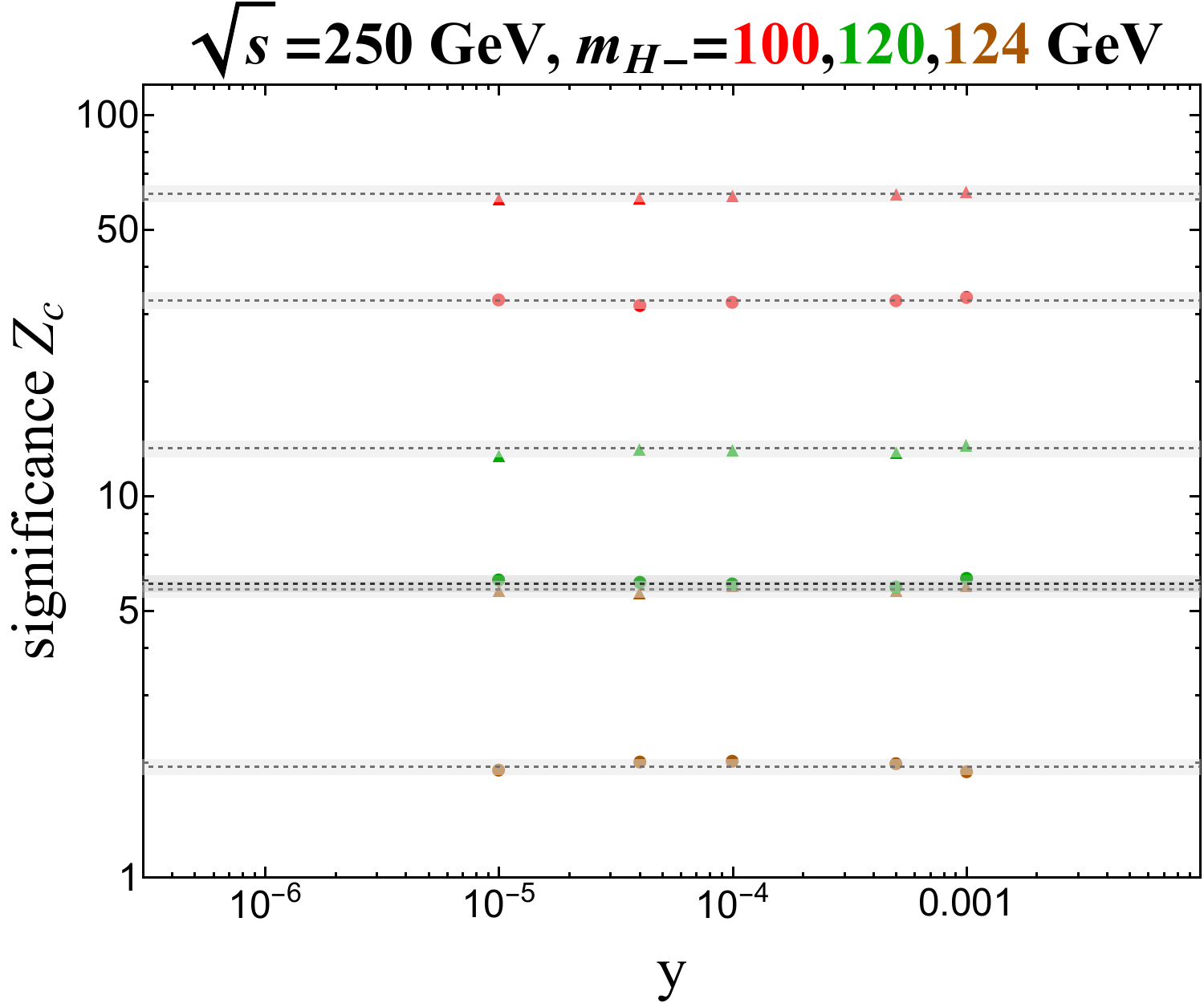}	
	\caption{
    Statistical significance of the process $e^+ e^- \rightarrow \bar{\nu}_l l^- \bar{d}u$ 
    with $l=e,~\mu$ at $\sqrt{s}=250~{\rm{GeV}}$ for double-sided reconstruction(left) and the single-sided reconstruction using $m_{jj}$ channel (right) after event selection, respectively. 
    Luminosity of Higgs factory is assumed to be $300~(3000)~{\rm fb^{-1}}$.
    The dashed line shows the average significance over the results obtained with different Yukawa couplings, 
    while the gray band denotes the potential uncertainty caused by MC simulation which is $5\%$ displayed here. 
    Near the kinematic threshold, the double-on-shell resonance enhancement still leads to a significance exceeding the $5\sigma$ discovery level. 
    The significance is calculated separately in each of the $N_{\rm bin}$=20 bins and combined in quadrature over all bins.
    }
	\label{fig_Z}%
\end{figure}

\begin{figure}[!htb]
	\centering 
	\includegraphics[width=0.45\textwidth, angle=0]{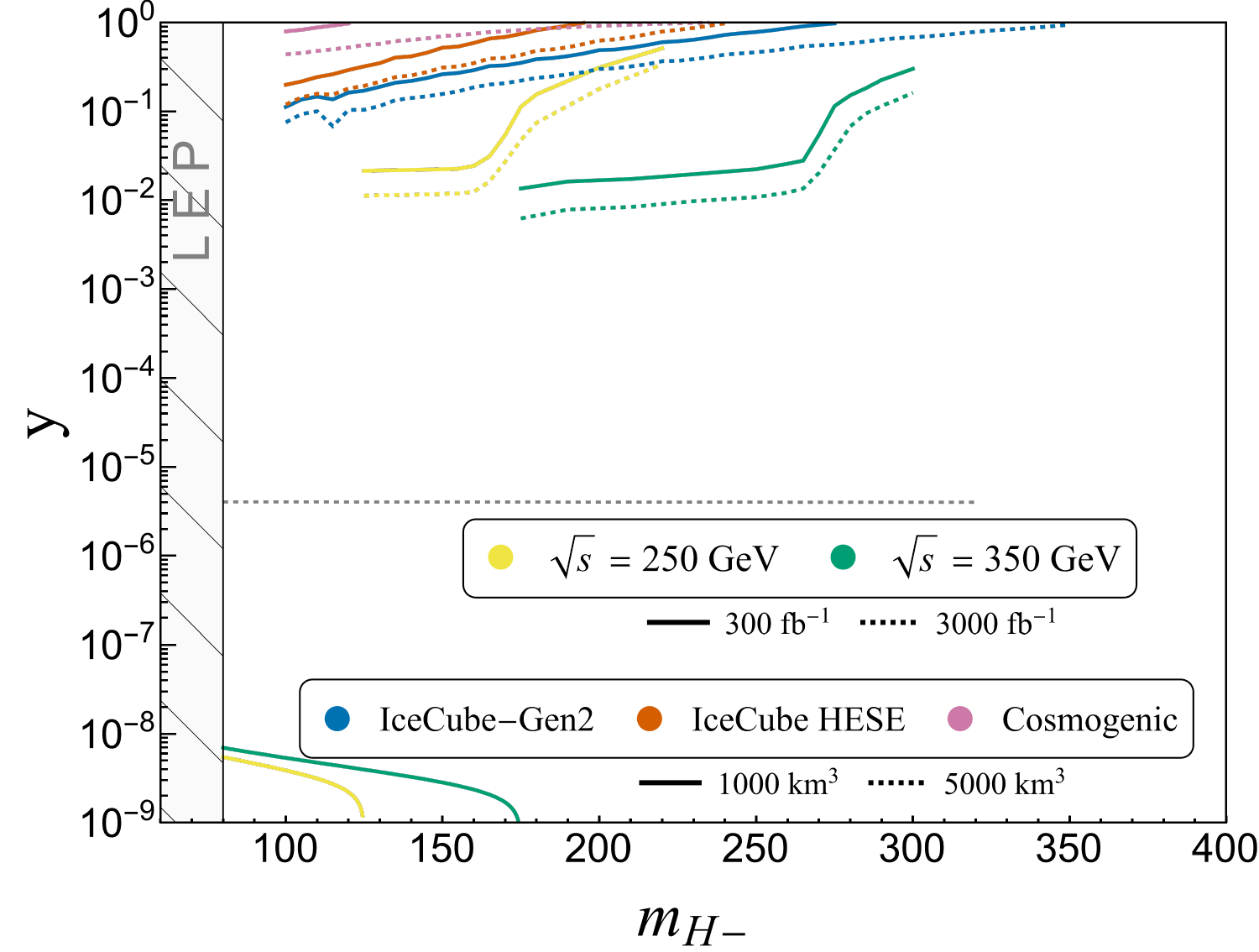}
	\caption{
	$5\sigma$ discovery sensitivity of the charged Higgs boson in Model I at the Higgs factory and neutrino telescope.
    Luminosity of Higgs factory is assumed to be $300~{\rm fb^{-1}}$ (solid line) and $3000~{\rm fb^{-1}}$ (dashed line).
    The light gray hatched area is excluded by the LEP~\cite{L3:2003jyb,ALEPH:2013htx}. 
    The discovery criteria differ between the pair on-shell region and the other regions, as described in the main text. 
 The line for $\sqrt{s}=250$ GeV with $m_{H^-}<125$ GeV  and the line for $\sqrt{s}=250$ GeV with $m_{H^-}<175$ GeV correspond to
 the lower bound that charged Higgs bosons decay inside the detector of Higgs factory.
 The parameter space below the dashed line indicates a break-down due to the small Yukawa coupling exceeds the precision required by the MC simulation.
 The neutrino telescope sensitivity curves are the same as the corresponding curves shown in Fig.~\ref{fig:flux_sensitivity}. 
 } 
	\label{fig_senCol}%
\end{figure}

\begin{figure}[!htb]
	\centering 
	\includegraphics[width=0.45\textwidth, angle=0]{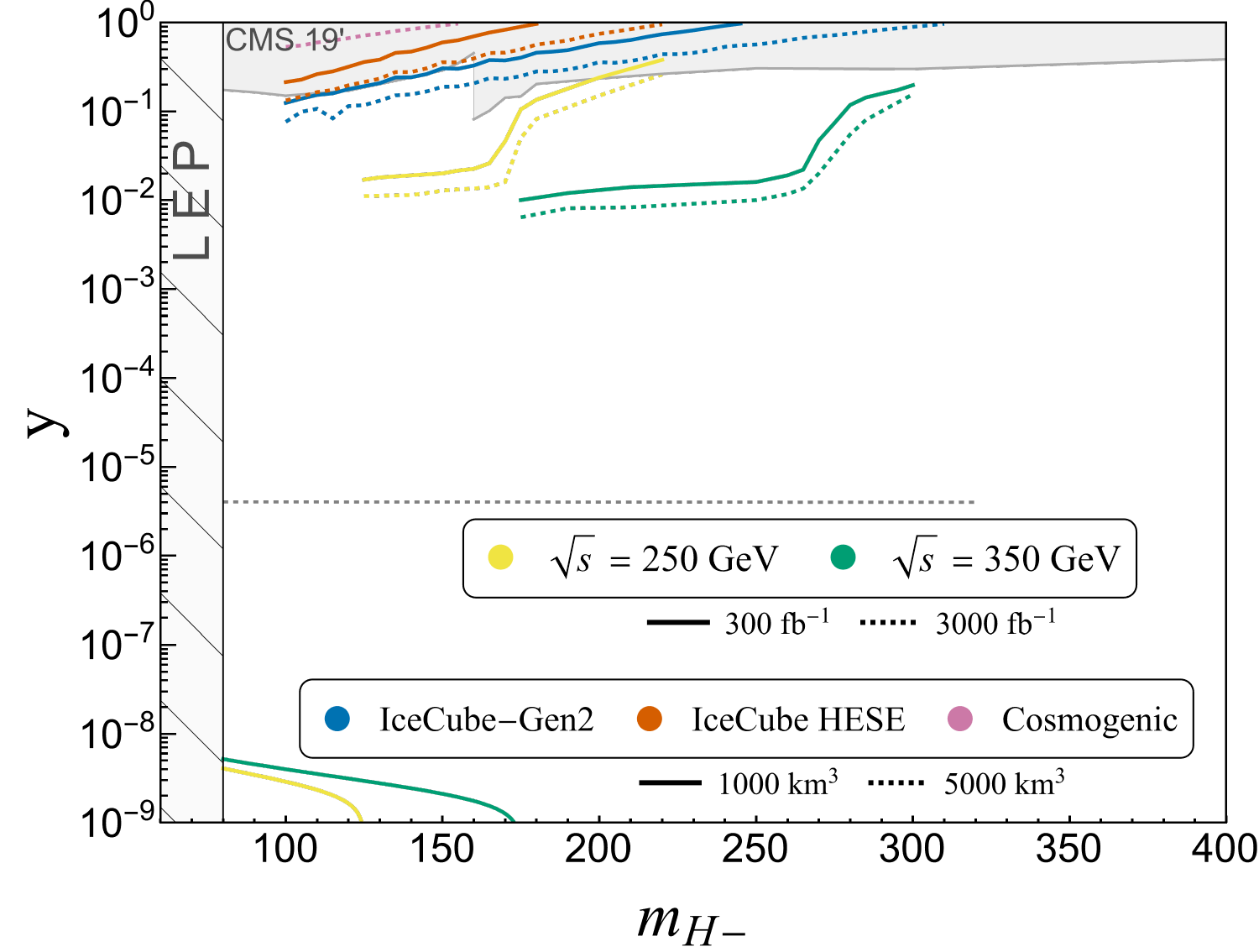}
	\caption{
    $5\sigma$  discovery sensitivity of the charged Higgs boson in Model II at the Higgs factory and neutrino telescope. 
    The light gray shaded region is excluded from the search for $\tau\nu$ final-state decays at the LHC~\cite{CMS:2019bfg}.
    Other settings are the same as in Fig.~\ref{fig_senCol}. 
     } 
	\label{fig_senCol2}%
\end{figure}

The statistical significance of the signal remains stable across a wide range of Yukawa coupling values, 
as long as the charged Higgs bosons decay promptly within the detector volume, as can be seen in Fig.~\ref{fig_Z} which
illustrates the significance for both the double-sided and the single-sided reconstruction.
We find that the resulting signal significances always exceed 5$\sigma$ for the double sided reconstruction strategy, as illustrated in Fig.~\ref{fig_Z}. 
In addition to basic cuts, Eq.~\eqref{basicCut1} and \eqref{basicCut2}, we also add selection cut  Eq.~\eqref{selCut1} to exclude backgrounds 
and stabilize signals.
We have verified that the same conclusion holds for $\sqrt{s}=350~{\rm{GeV}}$.
This observation allows us to define the discovery sensitivity through the process considered here in a model-independent manner by requiring that the charged Higgs bosons are produced on-shell and decay inside the detector,  which corresponds to 
the line for $\sqrt{s}=250$ GeV with $m_{H^-}<125$ GeV  and the line for $\sqrt{s}=350$ GeV with $m_{H^-}<175$ GeV in Fig.~\ref{fig_senCol} and Fig.~\ref{fig_senCol2}. 
As noted before, charged Higgs bosons decaying outside detector can be detected through charged tracks, but this detection scheme 
is beyond the consideration of the present work.
Therefore, charged Higgs boson with a mass below $\sqrt{s}/2$ can always be discovered in principle. 

For $m_{H^-}>\sqrt{s}/2$ which is outside the pair on-shell region, the double-sided reconstruction becomes less effective. 
In this case, we present the conventional $5\sigma$ discovery estimate based on $m_{jj}$ invariant mass distribution
in Fig.~\ref{fig_senCol} and Fig.~\ref{fig_senCol2}.
In addition to the basic cuts, Eq.~\eqref{basicCut1} and \eqref{basicCut2}, we also apply the selection cut  Eq.\eqref{selCut2}
to exclude backgrounds and stabilize signals. 
The solid lines for $\sqrt{s}=250$ GeV  and for $\sqrt{s}=350$ GeV start to rise rapidly around the kinematic threshold $m_{H^\pm}\simeq \sqrt{s}-m_W$.
For $m_{H^\pm}$ below this threshold, the final state can receive contributions from the single charged production processes, 
e.g. processes shown in Fig.~\ref{fig:singleRes},
in which both the charged Higgs boson and the $W$ boson are produced on-shell.
In this double-on-shell region, the signal rate is resonantly enhanced and is only mildly affected by the increase of $m_{H^\pm}$.
As a result, the $5\sigma$ sensitivity curves vary slowly in this region.
The resulting on-shell decay products are generally more energetic and hence lead to better event acceptance and reconstruction efficiency. 
Together, these effects improve the signal significance near the threshold region. 
Above this threshold, the double-on-shell configuration is no longer kinematically allowed.
The signal then involves at least one off-shell particle, and is suppressed by the off-shell propagator.
This leads to a rapid loss of sensitivity.

Our analysis is model-independent, and the charged-Higgs Yukawa coupling is treated as a free parameter. 
For comparison, specific models may predict a particular flavor structure. 
For example, in the Type-II 2HDM with $\tan\beta=1$, the leptonic charged-Higgs Yukawa couplings are of order $10^{-6}$, $10^{-4}$, and $10^{-2}$ for the first, second, and third generations, respectively. 
Taking these values as benchmarks, the collider sensitivities obtained in our analysis indicate substantial discovery potential.

\subsection{\label{Comparison of The Discovery Potential}Comparison of The Discovery Potential }

For comparison, we also include in  Fig.~\ref{fig_senCol} and Fig.~\ref{fig_senCol2} the sensitivity curves of the neutrino telescope shown in Fig.~\ref{fig:flux_sensitivity}. 
The astrophysical neutrino flux models used in this work are described in Sec.~\ref{Astrophysical Neutrino Flux Models}. 
The detector volume is assumed to be 1000 and $5000~\rm{km^3}$, respectively. Since the HESE and CF flux models lead to very similar sensitivities, only one of them is shown here. 

The two experimental setups lead to different sensitivity trends. For the case of collider, a sharp enhancement appears once the charged Higgs mass approaches the kinematic threshold, while for the neutrino telescope the sensitivity changes rather smoothly. This is because the telescope signal originates from single particle resonant scattering instead of threshold limited pair production.  
Collider searches exhibit superior sensitivity in the low charged Higgs mass region, where charged Higgs bosons can be efficiently produced on shell and decay promptly inside the detector. 
In contrast, neutrino telescopes can probe part of the heavier mass region with competitive sensitivity in particular in Model I,  
although achieving comparable sensitivity generally requires substantially larger detector volumes. 

For luminosity at future Higgs factories, we take 300 fb$^{-1}$ as an early benchmark and 3000 fb$^{-1}$ as the ultimate target. 
Increasing the integrated luminosity has only a modest impact on the discovery sensitivity in the pair on-shell region.
Outside of this region, the minimum Yukawa coupling that can be probed is expected to be reduced by roughly a factor of two in the high-luminosity case as shown in Fig.~\ref{fig_senCol} and Fig.~\ref{fig_senCol2} for both two simplified models. 

We conclude that under the currently proposed experimental configurations, collider searches remain more sensitive over most of the parameter space considered in this work. 
Future neutrino telescope with a very large detector volume is also competitive in discovering charged Higgs boson.

\section{\label{Summary}Summary}

In this work, we study the discovery potential of charged Higgs boson at future Higgs factories and neutrino telescopes and compare the sensitivities of them.
We find that the combination of collider and cosmic telescope searches offers a promising and complementary strategy for exploring charged Higgs bosons. 

The clean environment of the Higgs factories allows efficient reconstruction of charged Higgs signals. 
In the pair on-shell production region,  the signal rate is largely insensitive to the Yukawa couplings 
and a statistical significances above $5\sigma$ can always be achieved for a double-sided reconstruction,
as long as the charged Higgs boson decays  inside the detector volume. 
Outside the pair on-shell regime, the discovery sensitivity decreases as the masses of charged Higgs increase. 

Neutrino telescopes can provide comparable sensitivity in the heavy charged Higgs mass region, where direct production at colliders becomes difficult due to the limited center-of-mass energy.  We find that a very large detector volume, around $10^3$ km$^3$, is generally required 
for neutrino telescopes to achieve competitive sensitivity. 
With such a large detector volume, neutrino telescope will not only become a powerful tool for detecting ultra-high-energy neutrinos, 
but also a competitive instrument at the energy frontier.
This encourages building neutrino telescopes with very large detector volume keeping good sensitivity on very high energy muon track events 
which can be achieved, without a significant increase in cost, by increasing the distance between the  photomultipliers in neutrino telescope.

\section*{Acknowledgements}
\label{sec:acknowledgements}
W. Liao would like to thank Zhen Cao and Mingjun Chen for discussions on future neutrino telescopes.
 Y.-S. Lu thanks Jidong Du for helpful discussions on event reconstruction.
W. Liao is supported by National Natural Science Foundation of China under the grant No. 11875130.
Q.S. Yan's work is supported by the Natural Science Foundation of China under the Grants No. 11875260 and No. 12275143.

\appendix
\section{$\tau$ decay contributions}
\label{app:tau_decay}
In this appendix, we discuss the $\tau$ decay contributions to the muon-track and cascade events in details.

\begin{figure}[!htb]
	\centering 
	\includegraphics[width=0.45\textwidth, angle=0]{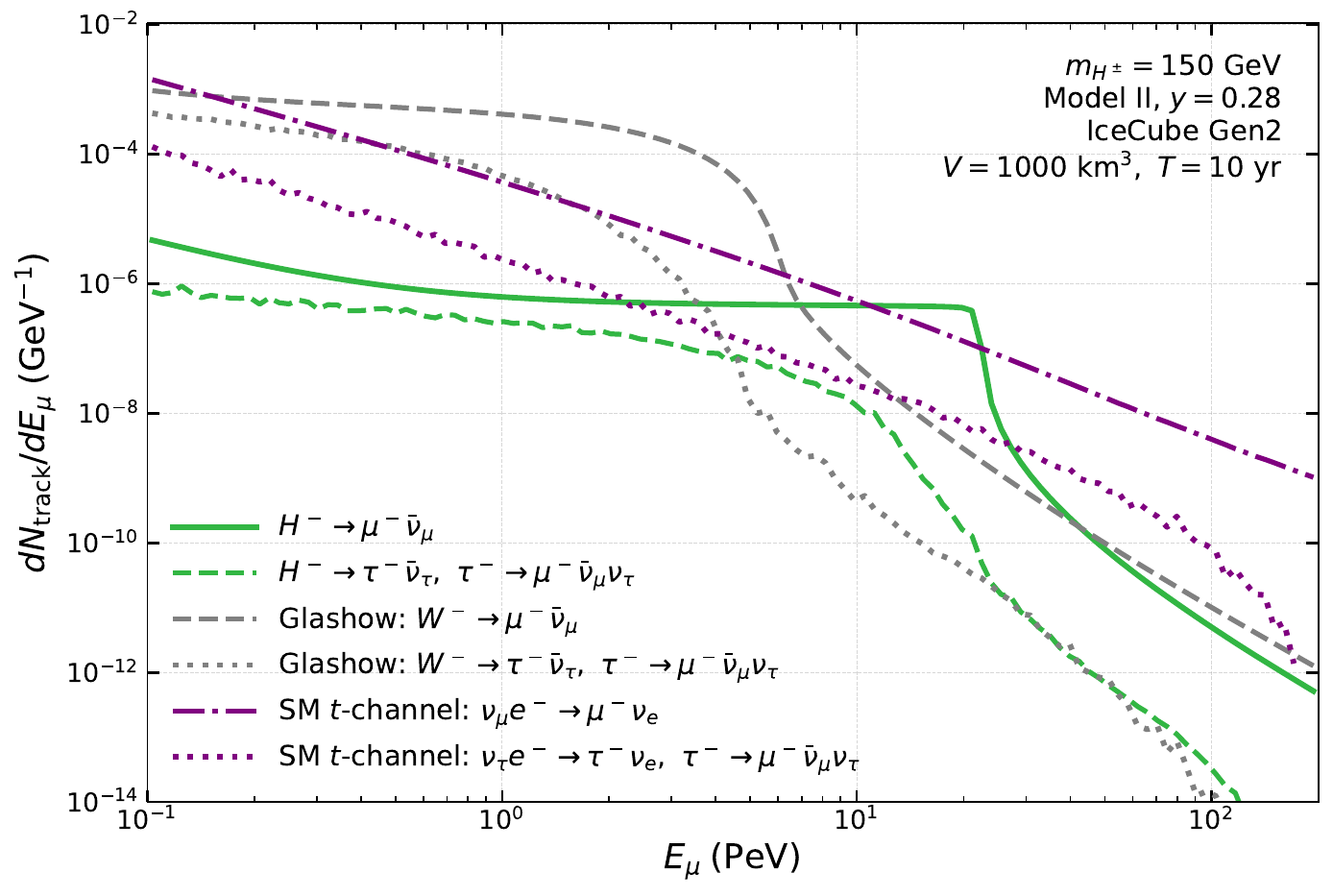}
	\caption{
    Differential muon-track event rates, $dN/dE_\mu$, as functions of the outgoing muon energy $E_\mu$.
    The green solid and dashed curves denote the charged Higgs contributions from the direct decay $H^-\rightarrow\mu^-\bar\nu_\mu$ and the $\tau$ decay channel $H^-\rightarrow\tau^-\bar\nu_\tau\rightarrow\mu^-\bar\nu_\mu\nu_\tau\bar\nu_\tau$, respectively.
    The gray dashed and dotted curves show the corresponding direct and $\tau$ decay contributions from the SM $s$-channel processes, while the purple dash-dotted and dotted curves denote those from the SM $t$-channel processes.
    For illustration, we take $m_{H^\pm}=150{\rm GeV}$, Model II with $y=0.28$, $V=1000~{\rm km}^3$, $T=10~{\rm yr}$ and the IceCube Gen2 flux.
     } 
     \label{fig:muon_track_all_processes_mH150}
\end{figure}

In addition to the direct muon-track events shown in Fig.~\ref{fig:event_rate}, muon tracks can also arise from the decays $\tau^\pm\rightarrow\mu^\pm\bar\nu\nu$ of $\tau$ produced in both the charged Higgs and SM processes.
The corresponding differential event rates are shown in Fig.~\ref{fig:muon_track_all_processes_mH150}.
The processes with direct muon production are calculated analytically, while those in which the muons originate from $\tau$ decays are evaluated numerically with \texttt{MadGraph5\underline{~}aMC@NLO}~\cite{Alwall:2014hca} and \texttt{Pythia8}~\cite{Sjostrand:2014zea}.
The comparison between the green solid and dashed curves shows that the muon-track event rate from $\tau^-\to\mu^-\bar\nu_\mu\nu_\tau$ is much smaller than that from the direct $H^-\rightarrow\mu^-\bar\nu_\mu$.
This suppression is due to the small branching ratio, ${\rm Br}(\tau^-\to\mu^-\bar\nu_\mu\nu_\tau)\simeq0.17$.
In addition, the two neutrinos produced in the $\tau$ decay carry away part of the $\tau$ energy, leading to a shift of the shoulder toward lower $E_\mu$.
As a result, the $\tau$ decay contribution is negligible in the energy range $10~{\rm PeV}\lesssim E_\mu\lesssim 20~{\rm PeV}$, where the charged Higgs signal is most relevant relative to the SM background.
A similar suppression also occurs for the SM contributions, as shown by the comparison between the gray dotted and dashed curves and between the purple dotted and dashed-dotted curves.
Therefore, these muon-track events arising from $\tau$ decays are neglected in the main analysis.

\begin{figure}[!htb]
	\centering 
	\includegraphics[width=0.45\textwidth, angle=0]{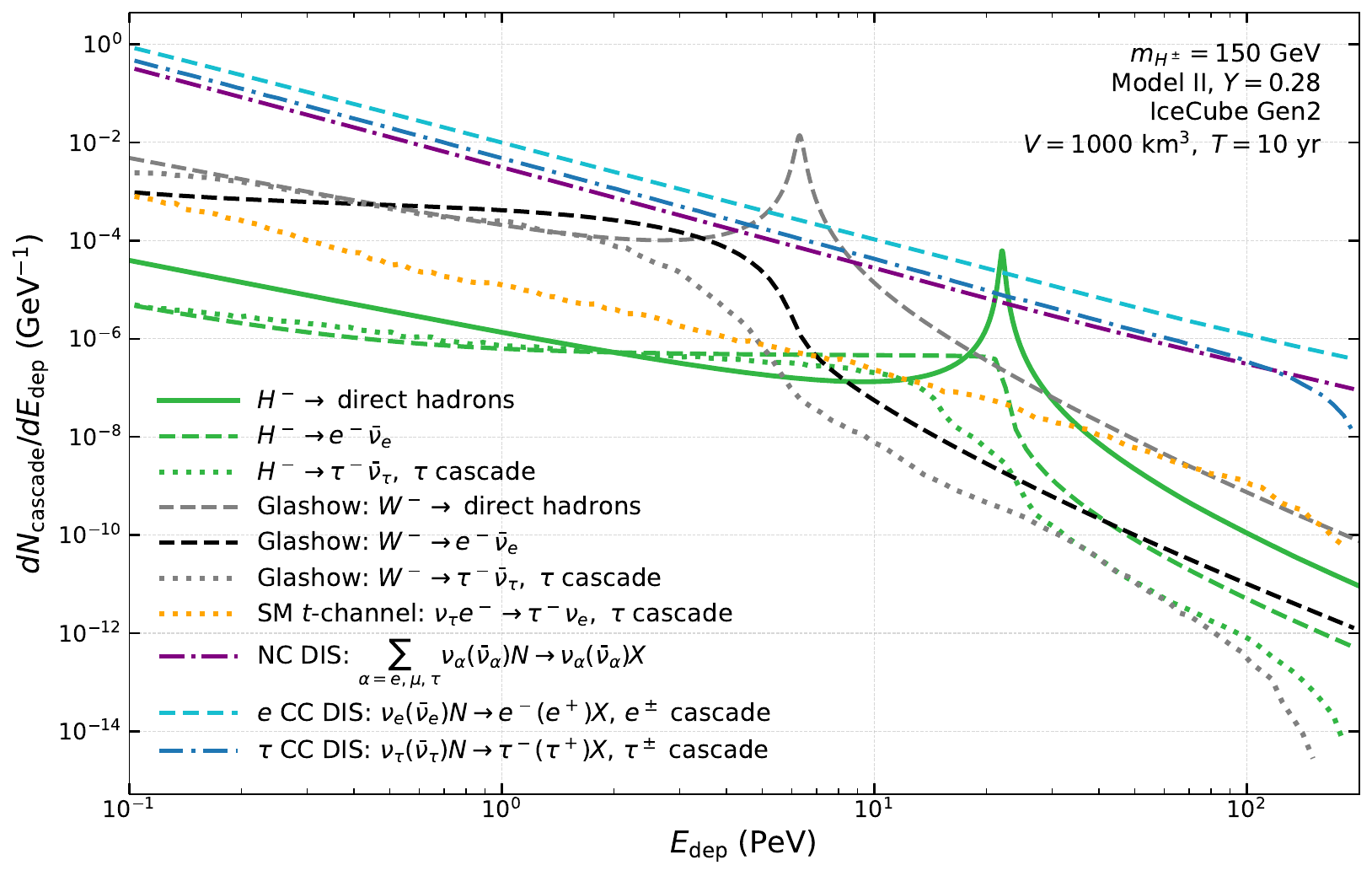}
	\caption{
    Differential rates for cascade events as functions of the energy decomposition $E_{\rm dep}$.
    The green solid, dashed, and dotted curves denote the charged Higgs contributions from the direct hadronic and electronic decay channels and from subsequent $\tau$ decays into cascade final states, respectively.
    The gray dashed, black dashed, and gray dotted curves show the corresponding hadronic, electronic, and $\tau$ decay contributions from the SM Glashow process.
    The purple dashed-dotted, cyan dashed, and blue dashed-dotted curves denote the NC DIS, $e$ CC DIS, and $\tau$ CC DIS backgrounds, respectively.
    The orange dash-dotted curve corresponds to $\nu_\tau e^-\rightarrow\tau^-\nu_e\rightarrow{\rm cascade}$.
  The explicit processes corresponding to these labels of lines are given in Fig.~\ref{fig:sigmaE_nu_model1}.
    The other settings are the same as those in Fig.~\ref{fig:muon_track_all_processes_mH150}.
}
     \label{fig:cascade_all_processes_mH150}
\end{figure}

A similar situation occurs in the cascade channel as shown in Fig.~\ref{fig:cascade_all_processes_mH150}.
In addition to the direct hadronic and electronic decays of the charged Higgs, the decays of the produced $\tau$ can also contribute to cascade events through $\tau^\pm\rightarrow e^\pm\bar\nu\nu$ and $\tau^\pm\rightarrow{\rm hadrons}+\nu$.
As shown by the green dashed and dotted curves, although the branching ratio ${\rm Br}(\tau\rightarrow{\rm cascade})\simeq0.83$, the neutrinos produced in the $\tau$ decay shift the shoulder toward lower $E_{\rm dep}$.
As a result, the $\tau$ decay contribution is much smaller near the direct hadronic resonance peak.
Moreover, in the lower panel of Fig.~\ref{fig:event_rate_cascade}, we conservatively assumed that the $\tau$ channel has the same deposited-energy distribution as the directly produced electronic cascade.
Even with this approximation, the sum of the electronic and $\tau$ contributions remains small near the hadronic resonance peak.
We therefore neglect these signals in the main analysis.
The relevant SM backgrounds near the charged Higgs resonance peak are the hadronic Glashow, NC DIS, $e$ CC DIS, and $\tau$ CC DIS channels as shown in Fig.~\ref{fig:cascade_all_processes_mH150}.
As shown by the black dashed, the gray dotted, and the orange dotted curves in Fig.~\ref{fig:cascade_all_processes_mH150}, the contributions from the leptonic Glashow processes and the SM $t$-channel process are much smaller than that from the $\tau$ CC DIS process.
We therefore retain only the $\tau$ CC DIS contribution and neglect the other $\tau$-related backgrounds.

Although the $\tau$-decay contribution can be evaluated event by event with \texttt{Pythia}, 
such a treatment is not well suited for simple analytic estimates. 
We therefore consider a semi-analytic treatment of the $\tau$ channel, whose validity is verified by comparison with the \texttt{Pythia} prediction. 

\textbf{Average-energy-fraction approximation}
The $\tau$-decay energy redistribution is approximated using the average visible-energy fraction, providing a simple semi-analytic description of the deposited-energy spectrum. 

We model the visible-energy spectrum through normalized energy-splitting kernels. 
For the resonant processes, $\bar\nu_e e^-\to W^-/H^-\to\tau^-\bar\nu_\tau$,
we define
\begin{align}
    x=\frac{E_\tau}{E_\nu},
    \qquad
    f_\tau(x)=3(1-x)^2,
    \qquad 0\leq x\leq1,
\end{align}
such that~\cite{Zas:2005zz}
\begin{align}
    \int_0^1 f_\tau(x)\,dx=1,
    \qquad
    \langle x\rangle=\int_0^1x f_\tau(x)\,dx=\frac14.
\end{align}
For simplicity, we take the daughter $\tau$ energy distribution in the charged-Higgs channel to be identical to that adopted for the SM $W$-resonance channel.

The subsequent decay is described by $z=\frac{E_{\rm obs}}{E_\tau}$,
where \(E_{\rm obs}=E_\mu\) for track events and \(E_{\rm obs}=E_{\rm dep}\) for cascade events. The normalized decay distributions for the individual modes are approximated by Beta distributions,
\[
p_i(z)=
\frac{z^{a_i-1}(1-z)^{b_i-1}}{B(a_i,b_i)},
\qquad 0<z<1,
\]
where
\[
B(a,b)
=
\int_0^1 t^{a-1}(1-t)^{b-1}\,dt
=
\frac{\Gamma(a)\Gamma(b)}{\Gamma(a+b)}.
\]
The shape parameters are chosen as
\[
a_i=\kappa\langle z\rangle_i,
\qquad
b_i=\kappa\bigl(1-\langle z\rangle_i\bigr),
\qquad
\kappa=8.
\]
This construction ensures
\[
\int_0^1p_i(z)\,dz=1,
\qquad
\int_0^1z\,p_i(z)\,dz=\langle z\rangle_i.
\]
In the cascade channel, the electronic and hadronic decay modes are treated separately, with mean visible-energy fractions $\langle z\rangle_e = 1/3$ and $\langle z\rangle_{\rm had} = 2/3$, weighted by their corresponding branching fractions. 
In the track channel, the contribution from $\tau^-\to\mu^-\bar\nu_\mu\nu_\tau$ is included using a continuous kernel with $\langle z\rangle_\mu = 1/3$.
The distributions implemented numerically are
\begin{align}
p_e(z)=p_\mu(z)
=
\frac{z^{5/3}(1-z)^{13/3}}
     {B(8/3,16/3)},
\end{align}
and
\begin{align}
p_{\rm had}(z)
=
\frac{z^{13/3}(1-z)^{5/3}}
     {B(16/3,8/3)}.
\end{align}
The branching-fraction-weighted decay-response kernels are defined separately for cascade and track events as
\begin{align}
    D_\tau^{\rm cas}(z)
    &={\rm Br}(\tau\to e\nu\nu)\, p_e(z)+
    {\rm Br}(\tau\to{\rm hadrons}+\nu) \,
    p_{\rm had}(z) , \\
    D_\tau^{\rm tr}(z)
    &={\rm Br}(\tau\to\mu\nu\nu)\,p_\mu(z),
\end{align}
It is normalized to 0.826 for cascade events and 0.174 for track events.
Equivalently, these kernels approximate the corresponding restricted differential decay widths,
\begin{equation}
    D_{\tau}^i (z) = \sum_{i = \rm cas,tr}
    \frac1{\Gamma_{\tau}} \frac{d\Gamma_{\tau}}{dz} .
\end{equation}

The energy-response kernel accounting for both the resonant $\tau$-production kinematics and the subsequent $\tau$-decay is obtained by the convolution 
\begin{align}
    K(u) = \int_u^1 \frac{dx}{x} f_\tau (x) D_\tau \left( \frac{u}{x} \right), 
\end{align}
where $u=E_{\rm obs} / E_{\nu}$.
The differential cross-section for the resonant process is
$\frac{d\sigma_s(E_\nu,E_{\rm obs})}{dE_{\rm obs}} \simeq \sigma_s(E_\nu,E_{\rm obs})
\frac1{E_{\nu}} 
K \left(\frac{E_{\rm obs}}{E_{\nu}} \right)$.

For the non-resonant SM \(W\) \(t\)-channel process $\nu_\tau e^-\to\tau^-\nu_e$,
the $\tau$-energy distribution is instead determined directly from the differential scattering cross section,
\begin{align}
    \frac{d\sigma_t}{dx} =
    \frac{G_F^2(s-m_\tau^2)}{\pi}\frac{m_W^4}{\left[m_W^2+s(1-x)\right]^2},
\end{align}
and 
\begin{align}
    \frac{d\sigma_t(E_\nu,E_{\rm obs})}{dE_{\rm obs}} 
    = \frac1{E_{\nu}} \int_{{\rm max}(u, m_{\tau}^2/s)}^1 
    \frac{dx}{x} \, \frac{d\sigma_t}{dx}
    D_\tau \left( \frac{u}{x} \right) .
\end{align}
Thus, unlike the resonant $W$ contributions, the $W$ $t$-channel contribution does not use $f_\tau(x)=3(1-x)^2$.

For $\tau$ CC DIS, the visible fraction also contains the primary hadronic deposition, and is written as $u=y+(1-y)z$, where $y$ is the DIS inelasticity~\cite{Bustamante:2017xuy}. 
We therefore convolve the $\tau$-decay kernel with an effective normalized inelasticity distribution
$p_{\rm DIS}(y) = 4(1-y)^3$, corresponding to $\langle y\rangle =0.2$~\cite{Zas:2005zz}. 
The relevant kernel for the CC DIS process becomes
\begin{align}
    K(u) = \int_0^u \frac{dy}{1-y} p_{\rm DIS}(y) D_\tau \left( \frac{u-y}{1-y} \right). 
\end{align} 
Similarly, the differential cross-section for the CC DIS process is
$\frac{d\sigma_{\rm CC}(E_\nu,E_{\rm obs})}{dE_{\rm obs}} \simeq \sigma_{\rm CC}(E_\nu,E_{\rm obs})
\frac1{E_{\nu}} 
K \left(\frac{E_{\rm obs}}{E_{\nu}} \right)$.

For a given final state, the differential event rate is given by
\begin{equation} \label{eq:dNdE}
\frac{dN_f}{dE_{\rm obs}}
=
N_{\rm taget} T_0
\int_0^{4\pi} d\Omega
\int dE_\nu\,
\frac{d\sigma_f(E_\nu,E_{\rm obs})}{dE_{\rm obs}}\,
\Phi_{\bar{\nu}_e}(E_\nu).
\end{equation}
where $f=s,~t,~{\rm CC}$ represents the resonant, $t$-channel and $\tau$ CC DIS process, respectively.
$T_0$ and $N_{\rm taget}$ are the exposure time and the number of targets, respectively.

In summary, this approximation retains the continuous energy redistribution associated with both production and decay, while avoiding a full event-generator simulation of polarized $\tau$ decays and detector response. 
The adopted kernels should be regarded as a phenomenological energy-response model rather than a fully differential calculation.

\begin{figure}[!htb]
	\centering 
	\includegraphics[width=0.45\textwidth, angle=0]{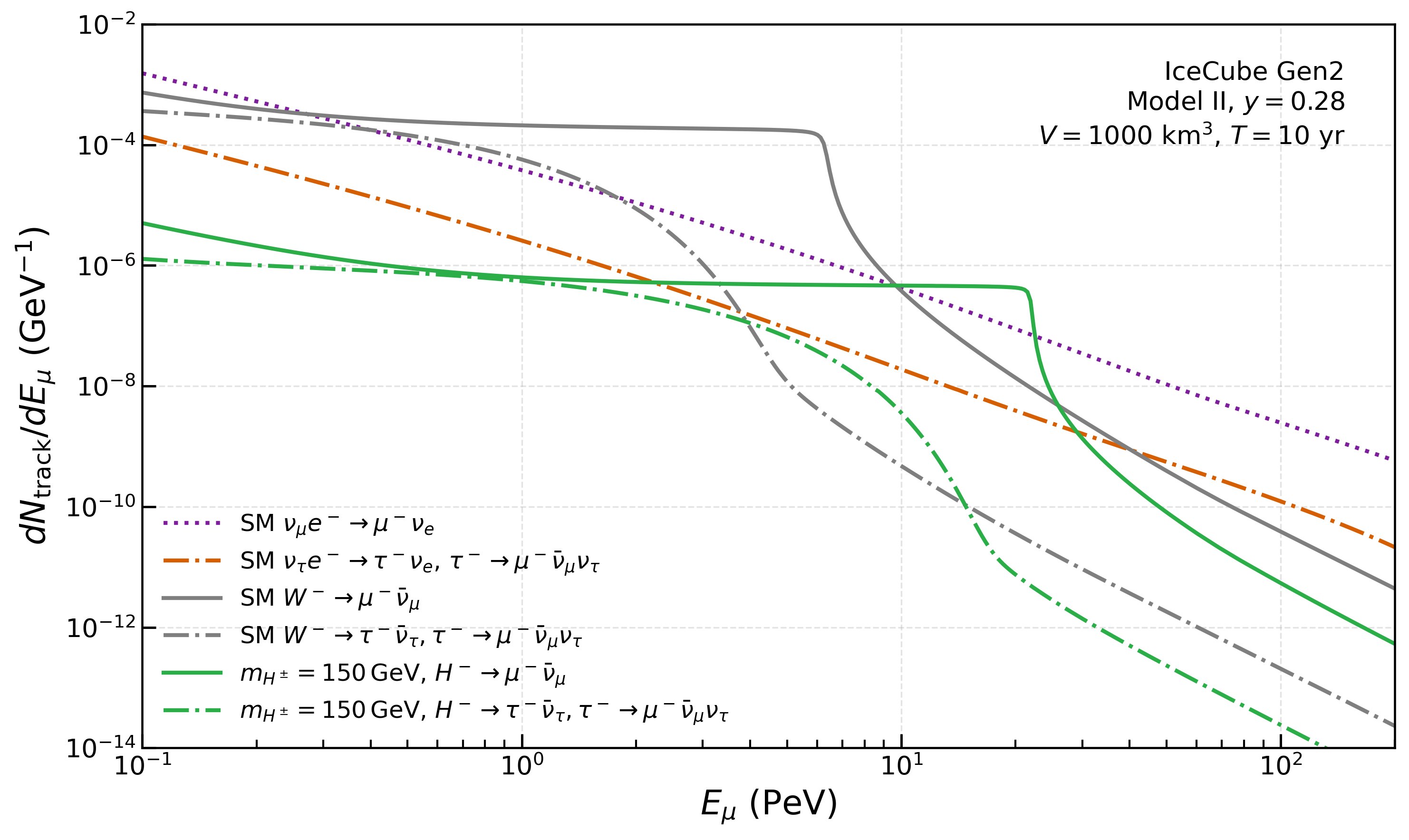} \\
    \includegraphics[width=0.45\textwidth, angle=0]{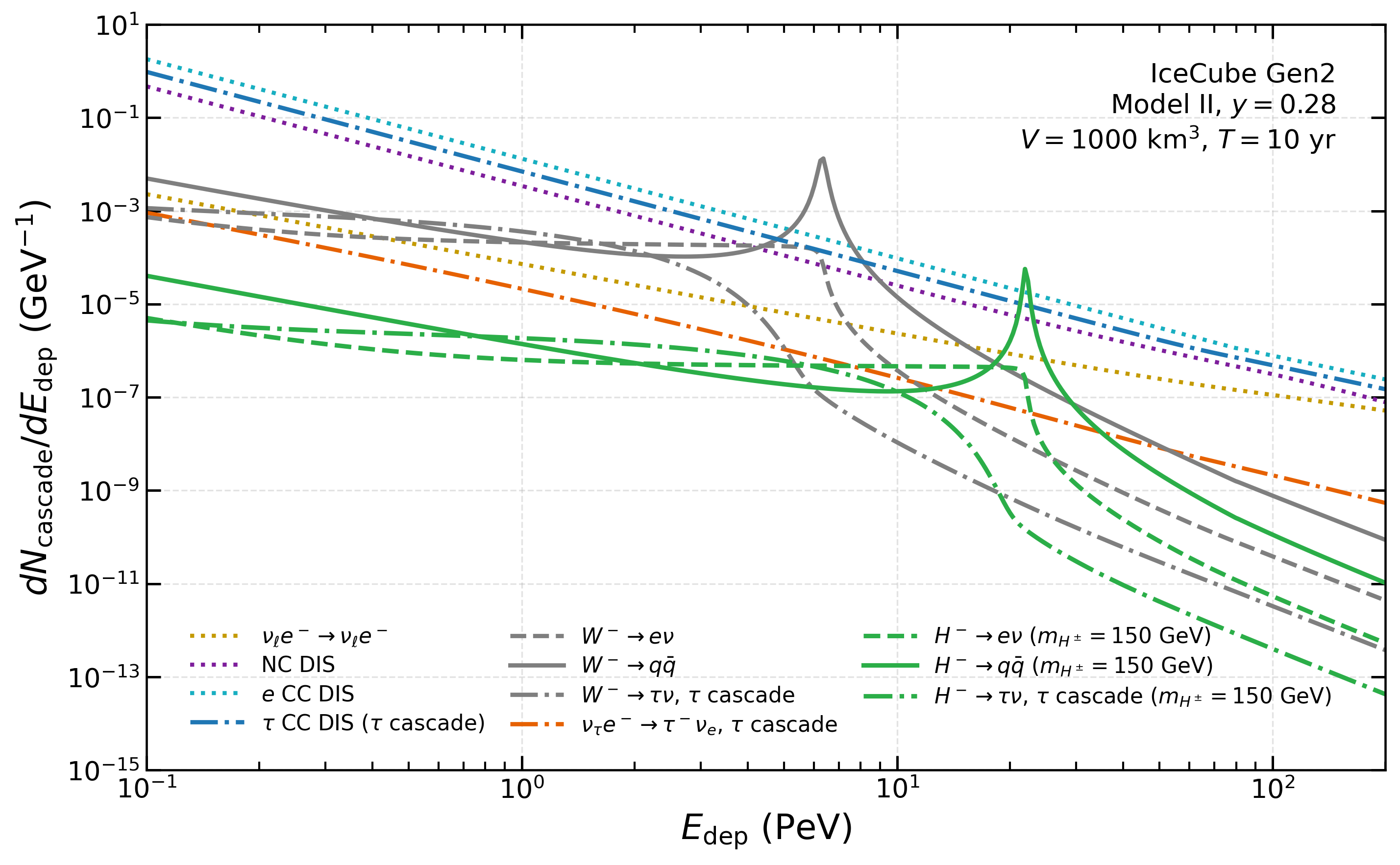}
	\caption{
    Differential rates for muon-track(upper panel) and cascade(lower panel) events as functions of the visible energy $E_{\rm obs}$, where $E_{\rm obs}=E_\mu$ for the muon-track events and $E_{\rm obs}=E_{\rm dep}$ for cascade events. 
    The colored curves show the individual process contributions, as labeled.
    The dotted curves correspond to the common contribution without $\tau$-induced events.
    The solid and dashed curves show the resonant contributions without $\tau$-induced events.
    The dash-dotted curves denote the contribution from $\tau$ decays using the average-energy-fraction approximation.
    The explicit processes corresponding to these SM labels are given in Fig.~\ref{fig:sigmaE_nu_model1}.
    For illustration, we take $m_{H^\pm}=150{\rm GeV}$, Model II with $y=0.28$, $V=1000~{\rm km}^3$, $T=10~{\rm yr}$ and the IceCube Gen2 flux.
    }
    \label{fig:dNdE_tau_average}
\end{figure}

The implementation of these procedures are illustrated in Fig.~\ref{fig:dNdE_tau_average}. 
The corresponding $\tau$-decay branching fractions are retained in the event-rate calculation.
We show the differential event spectra of individual process contributions as in Fig.~\ref{fig:muon_track_all_processes_mH150} and Fig.~\ref{fig:cascade_all_processes_mH150}.
The dash-dotted curves denote the contribution from $\tau$ decays using the average-energy-fraction approximation.
In the the case of track events, the comparison between the solid and dash-dotted curves shows that the $\tau$-decay contributions in both the $W$ background and the charged Higgs signal have a negligible impact on both the resonant peak and the overall spectrum.
In the case of cascade events, the only non-negligible $\tau$ contribution is from the $\tau$ CC DIS background process.

The average-energy-fraction approximation reproduces the full simulation accurately,
while it slightly overestimates the $\tau$ CC DIS background obtained with \texttt{Pythia}, as shown by the blue dash-dotted curve in Fig.~\ref{fig:cascade_all_processes_mH150}.

Using the kernels, we find that $\tau$ decay contribution produces only negligible changes in the $W$ and charged Higgs processes,
which double justified their omission in the main analysis.
The agreement between the Monte Carlo and semi-analytic treatments indicates that, for the observables and energy range considered here, the final differential rates are rarely affected by the detailed exclusive $\tau$-decay dynamics.
Hence, we can safely neglect these contributions without losing precision for the $W$ and charged Higgs signals in our analysis, while only accounting for the $\tau$ CC DIS background, as it provides a sufficiently accurate description of the $\tau$ channels. 

\nocite{*}

\bibliography{ref_chH}

@article{ATLAS:2012yve,
    author = "Aad, Georges and others",
    collaboration = "ATLAS",
    title = "{Observation of a new particle in the search for the Standard Model Higgs boson with the ATLAS detector at the LHC}",
    eprint = "1207.7214",
    archivePrefix = "arXiv",
    primaryClass = "hep-ex",
    reportNumber = "CERN-PH-EP-2012-218",
    doi = "10.1016/j.physletb.2012.08.020",
    journal = "Phys. Lett. B",
    volume = "716",
    pages = "1--29",
    year = "2012"
}

@article{CMS:2012qbp,
    author = "Chatrchyan, Serguei and others",
    collaboration = "CMS",
    title = "{Observation of a New Boson at a Mass of 125 GeV with the CMS Experiment at the LHC}",
    eprint = "1207.7235",
    archivePrefix = "arXiv",
    primaryClass = "hep-ex",
    reportNumber = "CMS-HIG-12-028, CERN-PH-EP-2012-220",
    doi = "10.1016/j.physletb.2012.08.021",
    journal = "Phys. Lett. B",
    volume = "716",
    pages = "30--61",
    year = "2012"
}

@article{Branco:2011iw,
    author = "Branco, G. C. and Ferreira, P. M. and Lavoura, L. and Rebelo, M. N. and Sher, Marc and Silva, Joao P.",
    title = "{Theory and phenomenology of two-Higgs-doublet models}",
    eprint = "1106.0034",
    archivePrefix = "arXiv",
    primaryClass = "hep-ph",
    doi = "10.1016/j.physrep.2012.02.002",
    journal = "Phys. Rept.",
    volume = "516",
    pages = "1--102",
    year = "2012"
}

@article{Babu:2022fje,
    author = "Babu, K. S. and Dev, P. S. Bhupal and Jana, Sudip",
    title = "{Probing neutrino mass models through resonances at neutrino telescopes}",
    eprint = "2202.06975",
    archivePrefix = "arXiv",
    primaryClass = "hep-ph",
    doi = "10.1142/S0217751X22300034",
    journal = "Int. J. Mod. Phys. A",
    volume = "37",
    number = "11n12",
    pages = "2230003",
    year = "2022"
}

@misc{Bai:2025pef,
    author = "Bai, Yang and Xie, Keping and Zhou, Bei",
    title = "{Large Neutrino ''Collider''}",
    eprint = "2510.13948",
    archivePrefix = "arXiv",
    primaryClass = "hep-ph",
    reportNumber = "FERMILAB-PUB-25-0731-T",
    year = "2025"
}

@article{CEPCStudyGroup:2018ghi,
    author = "Dong, Mingyi and others",
    editor = "Guimar{\~a}es da Costa, Jo{\~a}o Barreiro and others",
    collaboration = "CEPC Study Group",
    title = "{CEPC Conceptual Design Report: Volume 2 - Physics {\&} Detector}",
    eprint = "1811.10545",
    archivePrefix = "arXiv",
    primaryClass = "hep-ex",
    reportNumber = "IHEP-CEPC-DR-2018-02, IHEP-EP-2018-01, IHEP-TH-2018-01",
    journal = "Sci. China Phys. Mech. Astron.",
    volume = "63",
    number = "2",
    pages = "221001",
    year = "2020",
}

@article{FCC:2018evy,
  author        = "{FCC Collaboration}",
  title         = "{FCC-ee: The Lepton Collider: Future Circular Collider Conceptual Design Report Volume 2}",
  journal       = "Eur. Phys. J. ST",
  volume        = "228",
  number        = "2",
  pages         = "261--623",
  year          = "2019",
  doi           = "10.1140/epjst/e2019-900045-4",
}

@inproceedings{cheng2024physicspotentialcepcprepared,
    author = "Cheng, Huajie and others",
    collaboration = "CEPC Physics Study Group",
    title = "{The Physics potential of the CEPC. Prepared for the US Snowmass Community Planning Exercise (Snowmass 2021)}",
    booktitle = "{Snowmass 2021}",
    eprint = "2205.08553",
    archivePrefix = "arXiv",
    primaryClass = "hep-ph",
    month = "5",
    year = "2022"
}

@inproceedings{agapov2022futurecircularleptoncollider,
    author = "Agapov, I. and others",
    title = "{Future Circular Lepton Collider FCC-ee: Overview and Status}",
    booktitle = "{Snowmass 2021}",
    eprint = "2203.08310",
    archivePrefix = "arXiv",
    primaryClass = "physics.acc-ph",
    reportNumber = "FERMILAB-CONF-22-177-AD",
    month = "3",
    year = "2022"
}

@article{CMS:2021wlt,
    author = "Sirunyan, Albert M and others",
    collaboration = "CMS",
    title = "{Search for charged Higgs bosons produced in vector boson fusion processes and decaying into vector boson pairs in proton{\textendash}proton collisions at $\sqrt{s} = 13\,{\text {TeV}} $}",
    eprint = "2104.04762",
    archivePrefix = "arXiv",
    primaryClass = "hep-ex",
    reportNumber = "CMS-HIG-20-017, CERN-EP-2021-045",
    doi = "10.1140/epjc/s10052-021-09472-3",
    journal = "Eur. Phys. J. C",
    volume = "81",
    number = "8",
    pages = "723",
    year = "2021"
}

@article{ATLAS:2024txt,
    author = "Aad, Georges and others",
    collaboration = "ATLAS",
    title = "{Combination of searches for singly and doubly charged Higgs bosons produced via vector-boson fusion in proton{\textendash}proton collisions at s=13 TeV with the ATLAS detector}",
    eprint = "2407.10798",
    archivePrefix = "arXiv",
    primaryClass = "hep-ex",
    reportNumber = "CERN-EP-2024-189",
    doi = "10.1016/j.physletb.2024.139137",
    journal = "Phys. Lett. B",
    volume = "860",
    pages = "139137",
    year = "2025"
}

@article{CMS:2022syk,
    author = "Tumasyan, Armen and others",
    collaboration = "CMS",
    title = "{Search for direct pair production of supersymmetric partners of $\tau$ leptons in the final state with two hadronically decaying $\tau$ leptons and missing transverse momentum in proton-proton collisions at $\sqrt{s}$ = 13 TeV}",
    eprint = "2207.02254",
    archivePrefix = "arXiv",
    primaryClass = "hep-ex",
    reportNumber = "CMS-SUS-21-001, CERN-EP-2022-032",
    doi = "10.1103/PhysRevD.108.012011",
    journal = "Phys. Rev. D",
    volume = "108",
    number = "1",
    pages = "012011",
    year = "2023"
}

@article{ATLAS:2019gti,
    author = "Aad, Georges and others",
    collaboration = "ATLAS",
    title = "{Search for direct stau production in events with two hadronic $\tau$-leptons in $\sqrt{s} = 13$ TeV $pp$ collisions with the ATLAS detector}",
    eprint = "1911.06660",
    archivePrefix = "arXiv",
    primaryClass = "hep-ex",
    reportNumber = "CERN-EP-2019-191",
    doi = "10.1103/PhysRevD.101.032009",
    journal = "Phys. Rev. D",
    volume = "101",
    number = "3",
    pages = "032009",
    year = "2020"
}

@article{ATLAS:2021upq,
    author = "Aad, Georges and others",
    collaboration = "ATLAS",
    title = "{Search for charged Higgs bosons decaying into a top quark and a bottom quark at $ \sqrt{\mathrm{s}} $ = 13 TeV with the ATLAS detector}",
    eprint = "2102.10076",
    archivePrefix = "arXiv",
    primaryClass = "hep-ex",
    reportNumber = "CERN-EP-2021-004",
    doi = "10.1007/JHEP06(2021)145",
    journal = "JHEP",
    number = "06",
    pages = "145",
    year = "2021"
}

@article{ATLAS:2023bzb,
    author = "Aad, Georges and others",
    collaboration = "ATLAS",
    title = "{Search for a light charged Higgs boson in $t \rightarrow H^{\pm}b$ decays, with $H^{\pm} \rightarrow cb$, in the lepton+jets final state in proton-proton collisions at $\sqrt{s}=13$ TeV with the ATLAS detector}",
    eprint = "2302.11739",
    archivePrefix = "arXiv",
    primaryClass = "hep-ex",
    reportNumber = "CERN-EP-2022-207",
    doi = "10.1007/JHEP09(2023)004",
    journal = "JHEP",
    number = "09",
    pages = "004",
    year = "2023"
}

@article{ATLAS:2024oqu,
    author = "Aad, Georges and others",
    collaboration = "ATLAS",
    title = "{Search for a light charged Higgs boson in $t \rightarrow H^{\pm } b$ decays, with $H^{\pm } \rightarrow cs$, in $pp$ collisions at $\sqrt{s}={13}\hbox { TeV}$ with the ATLAS detector}",
    eprint = "2407.10096",
    archivePrefix = "arXiv",
    primaryClass = "hep-ex",
    reportNumber = "CERN-EP-2024-185",
    doi = "10.1140/epjc/s10052-024-13715-4",
    journal = "Eur. Phys. J. C",
    volume = "85",
    number = "2",
    pages = "153",
    year = "2025"
}

@article{ATLAS:2024hya,
    author = "Aad, Georges and others",
    collaboration = "ATLAS",
    title = "{Search for charged Higgs bosons produced in top-quark decays or in association with top quarks and decaying via H{\ensuremath{\pm}}{\textrightarrow}{\ensuremath{\tau}}{\ensuremath{\pm}}{\ensuremath{\nu}}{\ensuremath{\tau}} in 13~TeV pp collisions with the ATLAS detector}",
    eprint = "2412.17584",
    archivePrefix = "arXiv",
    primaryClass = "hep-ex",
    reportNumber = "CERN-EP-2024-311",
    doi = "10.1103/PhysRevD.111.072006",
    journal = "Phys. Rev. D",
    volume = "111",
    number = "7",
    pages = "072006",
    year = "2025"
}

@article{ATLAS:2024rcu,
    author = "Aad, Georges and others",
    collaboration = "ATLAS",
    title = "{Search for a heavy charged Higgs boson decaying into a W boson and a Higgs boson in final states with leptons and b-jets in $ \sqrt{s} $ = 13 TeV pp collisions with the ATLAS detector}",
    eprint = "2411.03969",
    archivePrefix = "arXiv",
    primaryClass = "hep-ex",
    reportNumber = "CERN-EP-2024-290",
    doi = "10.1007/JHEP02(2025)143",
    journal = "JHEP",
    number = "02",
    pages = "143",
    year = "2025"
}

@article{CMS:2019idx,
    author = "Sirunyan, Albert M and others",
    collaboration = "CMS",
    title = "{Search for a light charged Higgs boson decaying to a W boson and a CP-odd Higgs boson in final states with e$\mu\mu$ or $\mu\mu\mu$ in proton-proton collisions at $\sqrt{s} =$ 13 TeV}",
    eprint = "1905.07453",
    archivePrefix = "arXiv",
    primaryClass = "hep-ex",
    reportNumber = "CMS-HIG-18-020, CERN-EP-2019-083",
    doi = "10.1103/PhysRevLett.123.131802",
    journal = "Phys. Rev. Lett.",
    volume = "123",
    number = "13",
    pages = "131802",
    year = "2019"
}

@article{CMS:2022jqc,
    author = "Tumasyan, Armen and others",
    collaboration = "CMS",
    title = "{Search for a charged Higgs boson decaying into a heavy neutral Higgs boson and a W boson in proton-proton collisions at $ \sqrt{s} $ = 13 TeV}",
    eprint = "2207.01046",
    archivePrefix = "arXiv",
    primaryClass = "hep-ex",
    reportNumber = "CMS-HIG-21-010, CERN-EP-2022-125",
    doi = "10.1007/JHEP09(2023)032",
    journal = "JHEP",
    number = "09",
    pages = "032",
    year = "2023"
}

@article{ParticleDataGroup:2024cfk,
    author = "Navas, S. and others",
    collaboration = "Particle Data Group",
    title = "{Review of particle physics}",
    doi = "10.1103/PhysRevD.110.030001",
    journal = "Phys. Rev. D",
    volume = "110",
    number = "3",
    pages = "030001",
    year = "2024"
}

@article{IceCube:2021rpz,
    author = "Aartsen, M. G. and others",
    collaboration = "IceCube",
    title = "{Detection of a particle shower at the Glashow resonance with IceCube}",
    eprint = "2110.15051",
    archivePrefix = "arXiv",
    primaryClass = "hep-ex",
    doi = "10.1038/s41586-021-03256-1",
    journal = "Nature",
    volume = "591",
    number = "7849",
    pages = "220--224",
    year = "2021",
    note = "[Erratum: Nature 592, E11 (2021)]"
}

@article{IceCube:2014stg,
    author = "Aartsen, M. G. and others",
    collaboration = "IceCube",
    title = "{Observation of High-Energy Astrophysical Neutrinos in Three Years of IceCube Data}",
    eprint = "1405.5303",
    archivePrefix = "arXiv",
    primaryClass = "astro-ph.HE",
    doi = "10.1103/PhysRevLett.113.101101",
    journal = "Phys. Rev. Lett.",
    volume = "113",
    pages = "101101",
    year = "2014"
}

@article{Schneider:2019ayi,
    author = "Schneider, Austin",
    collaboration = "IceCube",
    title = "{Characterization of the Astrophysical Diffuse Neutrino Flux with IceCube High-Energy Starting Events}",
    eprint = "1907.11266",
    archivePrefix = "arXiv",
    primaryClass = "astro-ph.HE",
    reportNumber = "PoS-ICRC2019-1004",
    doi = "10.22323/1.358.1004",
    journal = "PoS",
    volume = "ICRC2019",
    pages = "1004",
    year = "2020"
}

@article{KM3NeT:2025npi,
    author = "Aiello, S. and others",
    collaboration = "KM3NeT",
    title = "{Observation of an ultra-high-energy cosmic neutrino with KM3NeT}",
    doi = "10.1038/s41586-024-08543-1",
    journal = "Nature",
    volume = "638",
    number = "8050",
    pages = "376--382",
    year = "2025",
    note = "[Erratum: Nature 640, E3 (2025)]"
}

@article{Aartsen_2021,
    author = "Aartsen, M. G. and others",
    collaboration = "IceCube-Gen2",
    title = "{IceCube-Gen2: the window to the extreme Universe}",
    eprint = "2008.04323",
    archivePrefix = "arXiv",
    primaryClass = "astro-ph.HE",
    doi = "10.1088/1361-6471/abbd48",
    journal = "J. Phys. G",
    volume = "48",
    number = "6",
    pages = "060501",
    year = "2021"
}

@article{CHEN2026171374,
title = {HUNT: An ultra-large-scale neutrino astronomy telescope},
journal = {Nuclear Instruments and Methods in Physics Research Section A: Accelerators, Spectrometers, Detectors and Associated Equipment},
volume = {1086},
pages = {171374},
year = {2026},
issn = {0168-9002},
doi = {https://doi.org/10.1016/j.nima.2026.171374},
url = {https://www.sciencedirect.com/science/article/pii/S0168900226001002},
author = {Mingjun Chen},
collaboration = "on behalf of HUNT Collaboration"
}

@article{ye2024TRIDENT,
    author = "Ye, Z. P. and others",
    collaboration = "TRIDENT",
    title = "{A multi-cubic-kilometre neutrino telescope in the western Pacific Ocean}",
    eprint = "2207.04519",
    archivePrefix = "arXiv",
    primaryClass = "astro-ph.HE",
    doi = "10.1038/s41550-023-02087-6",
    journal = "Nature Astron.",
    volume = "7",
    number = "12",
    pages = "1497--1505",
    year = "2023"
}

@article{Aartsen_2017,
    author = "Aartsen, M. G. and others",
    collaboration = "IceCube",
    title = "{The IceCube Neutrino Observatory: Instrumentation and Online Systems}",
    eprint = "1612.05093",
    archivePrefix = "arXiv",
    primaryClass = "astro-ph.IM",
    doi = "10.1088/1748-0221/12/03/P03012",
    journal = "JINST",
    volume = "12",
    number = "03",
    pages = "P03012",
    year = "2017",
    note = "[Erratum: JINST 19, E05001 (2024)]"
}

@article{Dey_2021,
    author = "Dey, Ujjal Kumar and Nath, Newton and Sadhukhan, Soumya",
    title = "{Charged Higgs effects in IceCube: PeV events and NSIs}",
    eprint = "2010.05797",
    archivePrefix = "arXiv",
    primaryClass = "hep-ph",
    doi = "10.1007/JHEP09(2021)113",
    journal = "JHEP",
    number = "09",
    pages = "113",
    year = "2021"
}

@article{Ellwanger:2009dp,
    author = "Ellwanger, Ulrich and Hugonie, Cyril and Teixeira, Ana M.",
    title = "{The Next-to-Minimal Supersymmetric Standard Model}",
    eprint = "0910.1785",
    archivePrefix = "arXiv",
    primaryClass = "hep-ph",
    reportNumber = "LPT-ORSAY-09-76, CFTP-09-032, LPTA-09-066",
    doi = "10.1016/j.physrep.2010.07.001",
    journal = "Phys. Rept.",
    volume = "496",
    pages = "1--77",
    year = "2010"
}

@article{Babu:2019vff,
    author = "Babu, K. S. and Dev, P. S. and Jana, Sudip and Sui, Yicong",
    title = "{Zee-Burst: A New Probe of Neutrino Nonstandard Interactions at IceCube}",
    eprint = "1908.02779",
    archivePrefix = "arXiv",
    primaryClass = "hep-ph",
    reportNumber = "OSU-HEP-19-05",
    doi = "10.1103/PhysRevLett.124.041805",
    journal = "Phys. Rev. Lett.",
    volume = "124",
    number = "4",
    pages = "041805",
    year = "2020"
}

@article{IceCube:2013dkx,
    author = "Aartsen, M. G. and others",
    collaboration = "IceCube",
    title = "{Energy Reconstruction Methods in the IceCube Neutrino Telescope}",
    eprint = "1311.4767",
    archivePrefix = "arXiv",
    primaryClass = "physics.ins-det",
    doi = "10.1088/1748-0221/9/03/P03009",
    journal = "JINST",
    number = "9",
    pages = "P03009",
    year = "2014"
}

@article{IceCube:2020acn,
    author = "Aartsen, M. G. and others",
    collaboration = "IceCube",
    title = "{Characteristics of the diffuse astrophysical electron and tau neutrino flux with six years of IceCube high energy cascade data}",
    eprint = "2001.09520",
    archivePrefix = "arXiv",
    primaryClass = "astro-ph.HE",
    doi = "10.1103/PhysRevLett.125.121104",
    journal = "Phys. Rev. Lett.",
    volume = "125",
    number = "12",
    pages = "121104",
    year = "2020"
}

@article{PhysRev.118.316,
  title = {Resonant Scattering of Antineutrinos},
  author = {Glashow, Sheldon L.},
  journal = {Phys. Rev.},
  volume = {118},
  issue = {1},
  pages = {316--317},
  numpages = {0},
  year = {1960},
  month = {Apr},
  publisher = {American Physical Society},
  doi = {10.1103/PhysRev.118.316},
  url = {https://link.aps.org/doi/10.1103/PhysRev.118.316}
}

@article{Gandhi:1995tf,
    author = "Gandhi, Raj and Quigg, Chris and Reno, Mary Hall and Sarcevic, Ina",
    title = "{Ultrahigh-energy neutrino interactions}",
    eprint = "hep-ph/9512364",
    archivePrefix = "arXiv",
    reportNumber = "FERMILAB-PUB-95-221-T, CLNS-95-1357, MRI-PHY-16-95, UIOWA-95-06, AZPH-TH-95-15",
    doi = "10.1016/0927-6505(96)00008-4",
    journal = "Astropart. Phys.",
    volume = "5",
    pages = "81--110",
    year = "1996"
}

@article{Huang_2020,
    author = "Huang, Guo-yuan and Liu, Qinrui",
    title = "{Hunting the Glashow Resonance with PeV Neutrino Telescopes}",
    eprint = "1912.02976",
    archivePrefix = "arXiv",
    primaryClass = "hep-ph",
    doi = "10.1088/1475-7516/2020/03/005",
    journal = "JCAP",
    number = "03",
    pages = "005",
    year = "2020"
}

@article{abbasi2025,
    author = "Abbasi, R. and others",
    collaboration = "IceCube",
    title = "{Evidence for a Spectral Break or Curvature in the Spectrum of Astrophysical Neutrinos from 5 TeV--10 PeV}",
    eprint = "2507.22233",
    archivePrefix = "arXiv",
    primaryClass = "astro-ph.HE",
    doi = "10.1103/2gh9-d4q7",
    journal = "Phys. Rev. Lett.",
    volume = "136",
    pages = "121002",
    year = "2026"
}

@article{Groth:2021bub,
    author = "Groth, Kathrine M{\o}rch and G{\'e}nolini, Yoann and Ahlers, Markus",
    title = "{Improved Limits on Cosmogenic Fluxes from Ultra-High Energy Cosmic Rays}",
    doi = "10.22323/1.395.1005",
    journal = "PoS",
    volume = "ICRC2021",
    pages = "1005",
    year = "2021"
}

@article{Kampert:2016sqd,
    author = "Kampert, Karl-Heinz",
    collaboration = "Pierre Auger",
    title = "{Ultra-High Energy Cosmic Rays: Recent Results and Future Plans of Auger}",
    eprint = "1612.08188",
    archivePrefix = "arXiv",
    primaryClass = "astro-ph.HE",
    doi = "10.1063/1.4984858",
    journal = "AIP Conf. Proc.",
    volume = "1852",
    number = "1",
    pages = "040001",
    year = "2017"
}

@article{Alwall:2014hca,
    author = "Alwall, J. and others",
    title = "{The automated computation of tree-level and next-to-leading order differential cross sections, and their matching to parton shower simulations}",
    eprint = "1405.0301",
    archivePrefix = "arXiv",
    primaryClass = "hep-ph",
    reportNumber = "CERN-PH-TH-2014-064, CP3-14-18, LPN14-066, MCNET-14-09, ZU-TH-14-14",
    doi = "10.1007/JHEP07(2014)079",
    journal = "JHEP",
    number = "07",
    pages = "079",
    year = "2014"
}

@article{Sjostrand:2014zea,
    author = {Sj{\"o}strand, Torbj{\"o}rn and others},
    title = "{An introduction to PYTHIA 8.2}",
    eprint = "1410.3012",
    archivePrefix = "arXiv",
    primaryClass = "hep-ph",
    reportNumber = "LU-TP-14-36, MCNET-14-22, CERN-PH-TH-2014-190, FERMILAB-PUB-14-316-CD, DESY-14-178, SLAC-PUB-16122",
    doi = "10.1016/j.cpc.2015.01.024",
    journal = "Comput. Phys. Commun.",
    volume = "191",
    pages = "159--177",
    year = "2015"
}

@article{deFavereau:2013fsa,
    author = "de Favereau, J. and others",
    collaboration = "DELPHES 3",
    title = "{DELPHES 3, A modular framework for fast simulation of a generic collider experiment}",
    eprint = "1307.6346",
    archivePrefix = "arXiv",
    primaryClass = "hep-ex",
    doi = "10.1007/JHEP02(2014)057",
    journal = "JHEP",
    number = "02",
    pages = "057",
    year = "2014"
}

@article{Cowan:2010js,
    author = "Cowan, Glen and others",
    title = "{Asymptotic formulae for likelihood-based tests of new physics}",
    eprint = "1007.1727",
    archivePrefix = "arXiv",
    primaryClass = "physics.data-an",
    doi = "10.1140/epjc/s10052-011-1554-0",
    journal = "Eur. Phys. J. C",
    volume = "71",
    pages = "1554",
    year = "2011",
    note = "[Erratum: Eur.Phys.J.C 73, 2501 (2013)]"
}

@article{L3:2003jyb,
    author = "Achard, P. and others",
    collaboration = "L3",
    title = "{Search for charged Higgs bosons at LEP}",
    eprint = "hep-ex/0309056",
    archivePrefix = "arXiv",
    reportNumber = "CERN-EP-2003-054",
    doi = "10.1016/j.physletb.2003.09.057",
    journal = "Phys. Lett. B",
    volume = "575",
    pages = "208--220",
    year = "2003"
}

@article{ALEPH:2013htx,
    author = "Abbiendi, G. and others",
    collaboration = "ALEPH, DELPHI, L3, OPAL, LEP",
    title = "{Search for Charged Higgs bosons: Combined Results Using LEP Data}",
    eprint = "1301.6065",
    archivePrefix = "arXiv",
    primaryClass = "hep-ex",
    reportNumber = "CERN-PH-EP-2012-369",
    doi = "10.1140/epjc/s10052-013-2463-1",
    journal = "Eur. Phys. J. C",
    volume = "73",
    pages = "2463",
    year = "2013"
}

@article{CMS:2019bfg,
    author = "Sirunyan, Albert M and others",
    collaboration = "CMS",
    title = "{Search for charged Higgs bosons in the H$^{\pm}$ $\to$ $\tau^{\pm}\nu_\tau$ decay channel in proton-proton collisions at $\sqrt{s} =$ 13 TeV}",
    eprint = "1903.04560",
    archivePrefix = "arXiv",
    primaryClass = "hep-ex",
    reportNumber = "CMS-HIG-18-014, CERN-EP-2019-025",
    doi = "10.1007/JHEP07(2019)142",
    journal = "JHEP",
    number = "07",
    pages = "142",
    year = "2019"
}

@article{Zas:2005zz,
    author = "Zas, Enrique",
    title = "{Neutrino detection with inclined air showers}",
    eprint = "astro-ph/0504610",
    archivePrefix = "arXiv",
    doi = "10.1088/1367-2630/7/1/130",
    journal = "New J. Phys.",
    volume = "7",
    pages = "130",
    year = "2005"
}

@article{Bustamante:2017xuy,
    author = "Bustamante, Mauricio and Connolly, Amy",
    title = "{Extracting the Energy-Dependent Neutrino-Nucleon Cross Section above 10 TeV Using IceCube Showers}",
    eprint = "1711.11043",
    archivePrefix = "arXiv",
    primaryClass = "astro-ph.HE",
    doi = "10.1103/PhysRevLett.122.041101",
    journal = "Phys. Rev. Lett.",
    volume = "122",
    number = "4",
    pages = "041101",
    year = "2019"
}

\end{document}